\documentclass[a4paper,fleqn,usenatbib]{mnras}

\usepackage{newtxtext,newtxmath}

\usepackage[T1]{fontenc}
\usepackage{ae,aecompl}

\usepackage{tabularx}
\usepackage[normalem]{ulem}
\usepackage{multirow}
\usepackage{upgreek}
\usepackage{graphicx}	
\usepackage{amsmath}	
\usepackage{amssymb}	
\usepackage[multiple]{footmisc} 

\newcommand{\RN}[1]{%
	\textup{\uppercase\expandafter{\romannumeral#1}}%
}

\title[Eclipses of PSR J2051$-$0827]{Long Term Variability of a Black Widow's Eclipses -- A Decade of PSR J2051$-$0827}

\author[E. J. Polzin et al.]{
E. J. Polzin,$^{1}$\thanks{E-mail: elliott.polzin@manchester.ac.uk (EJP)}
R. P. Breton,$^{1}$
B. W. Stappers,$^{1}$
B. Bhattacharyya,$^{2}$
G. H. Janssen,$^{3,4}$
\newauthor
S. Os\l{}owski,$^{5}$
M. S. E. Roberts,$^{6,7}$
C. Sobey$^{8}$
\\
$^{1}$Jodrell Bank Centre for Astrophysics, School of Physics and Astronomy, The University of Manchester, Manchester M13 9PL, UK\\
$^{2}$National Centre for Radio Astrophysics, Tata Institute of Fundamental Research, Pune University, Pune 411007, India\\
$^{3}$ASTRON, the Netherlands Institute for Radio Astronomy, Oude Hoogeveensedijk 4, 7991 PD, Dwingeloo, the Netherlands\\
$^{4}$Department of Astrophysics/IMAPP, Radboud University, P. O. Box 9010, 6500 GL Nijmegen, The Netherlands\\
$^{5}$Swinburne University of Technology, PO Box 218, Hawthorn, VIC 3122, Australia\\
$^{6}$New York University Abu Dhabi, Saadiyat Island, Abu Dhabi, UAE\\
$^{7}$Eureka Scientific, Inc. Oakland, CA USA\\
$^{8}$CSIRO Astronomy and Space Science, PO Box 1130 Bentley, WA 6102, Australia
\\
}

\date{Accepted XXX. Received YYY; in original form ZZZ}

\pubyear{2018}

\begin{document}
\label{firstpage}
\pagerange{\pageref{firstpage}--\pageref{lastpage}}
\maketitle

\begin{abstract}
In this paper we report on $\sim10$\,years of observations of PSR J2051$-$0827, at radio frequencies in the range 110--4032\,MHz. We investigate the eclipse phenomena of this black widow pulsar using model fits of increased dispersion and scattering of the pulsed radio emission as it traverses the eclipse medium. These model fits reveal variability in dispersion features on timescales as short as the orbital period, and previously unknown trends on timescales of months--years. No clear patterns are found between the low-frequency eclipse widths, orbital period variations and trends in the intra-binary material density. Using polarisation calibrated observations we present the first available limits on the strength of magnetic fields within the eclipse region of this system; the average line of sight field is constrained to be $10^{-4}$\,G\,$\lesssim B_{||} \lesssim 10^2$\,G, while for the case of a field directed near-perpendicular to the line of sight we find $B_{\perp} \lesssim 0.3$\,G. Depolarisation of the linearly polarised pulses during the eclipse is detected and attributed to rapid rotation measure fluctuations of $\sigma_{\text{RM}} \gtrsim 100$\,rad\,m$^{-2}$ along, or across, the line of sights averaged over during a sub-integration. The results are considered in the context of eclipse mechanisms, and we find scattering and/or cyclotron absorption provide the most promising explanation, while dispersion smearing is conclusively ruled out. Finally, we estimate the mass loss rate from the companion to be $\dot{M}_{\text{C}} \sim 10^{-12}\,M_\odot$\,yr$^{-1}$, suggesting that the companion will not be fully evaporated on any reasonable timescale.
\end{abstract}




\section{Introduction}\label{sec: intro}
\begin{figure*}
	\includegraphics[width=.7\textwidth]{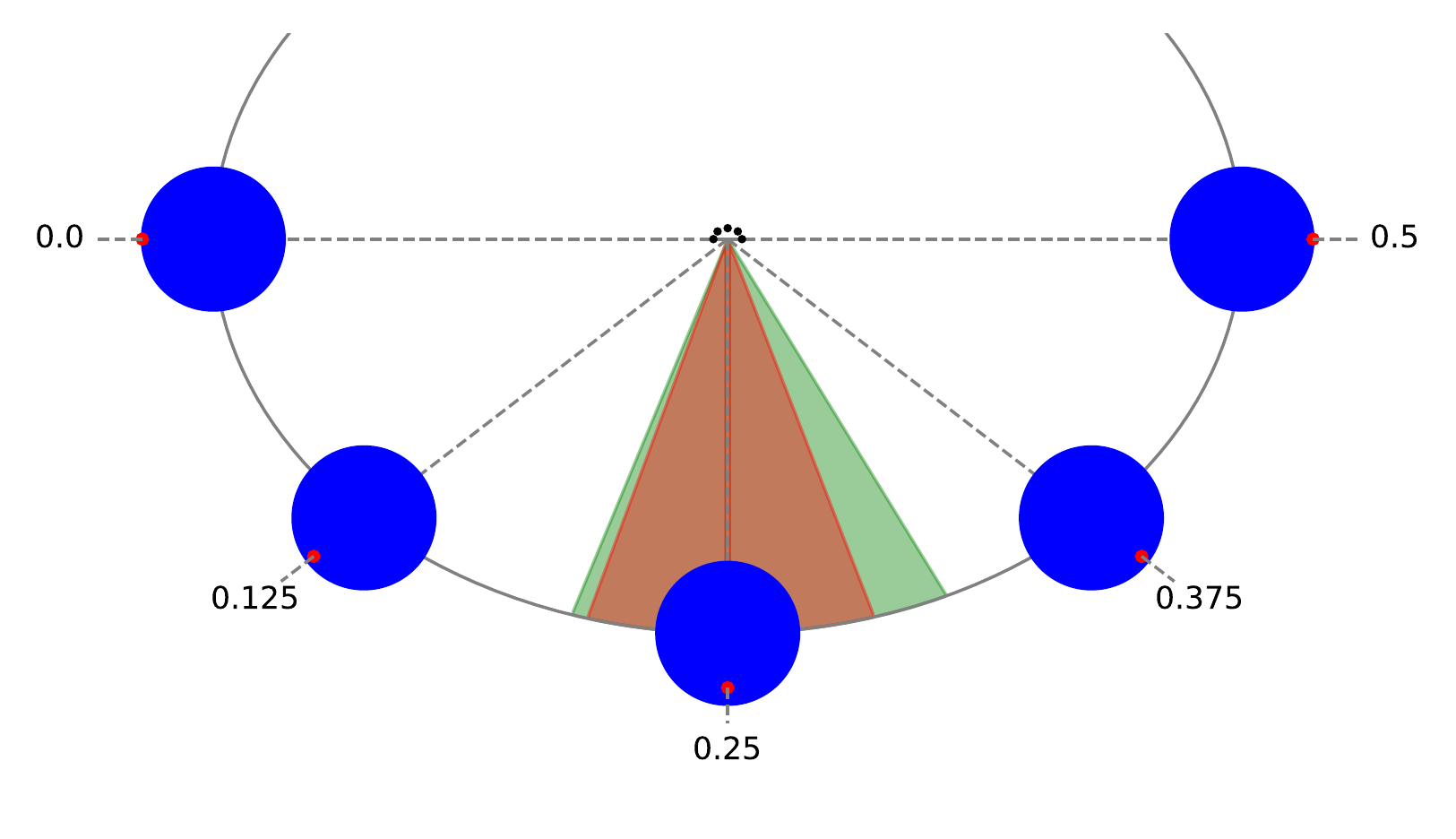}
	\caption{Expected projection of the PSR J2051$-$0827 system on the sky, assuming an inclination angle of $40^{\circ}$ \citep{svb+01}. The schematic shows 5 snapshots of the system at the labelled orbital phases, with the small black dots representing the pulsar, blue circles representing the companion star Roche lobe, and small red dots marking the centre point of the night-side of the companion star (i.e. the furthest point from the pulsar). The companion star and orbit are approximately to scale assuming the radio timing orbital parameters of \citet{svf+16}, pulsar and companion masses of $1.8\,M_{\odot}$ and $0.05\,M_{\odot}$, respectively, and companion Roche lobe radius of $0.15\,R_{\odot}$ as assumed in \citet{lvt+11}. The companion passes closest to our line of sight towards the pulsar -- directed out of the page -- at an orbital phase of 0.25, i.e. companion inferior conjunction. The shaded segments represent the approximate low-frequency eclipse orbital phases for 345\,MHz (red) and 149\,MHz (green) found from the observations in this paper.}
	\label{fig:geometry}
\end{figure*}

Black widows (BW) and redbacks (RB) -- collectively known as ``spider pulsars" -- are an intriguing subset of the pulsar population. All of the known spider pulsar systems in the Galactic disk\footnote{The case of systems located in globular clusters may be complicated due to tidal interactions.} comprise a millisecond pulsar (MSP) with a light-weight companion star in tight, near-circular orbits. The two types are separated observationally by the masses of their companions, with BW companion masses $\sim0.01$--$0.05 M_\odot$, and RB companions roughly a factor of ten more massive \citep{r11}. The majority of known systems show periodic eclipses of the radio emission centred approximately around inferior conjunction of the companion stars \citep[e.g.][]{fst88}, and the duration of the eclipses is such that the medium responsible can not be contained within the Roche lobes of the companions. Optical observations show the companions to be tidally locked, with the inner-face heated through irradiation from the pulsar wind \citep[e.g.][]{f+88}. The strong irradiation, and presence of material beyond the Roche lobes of the companions point towards the companion stars being ablated by the pulsar wind, constantly replenishing the eclipse medium.\\
The radio eclipses appear to be frequency dependent, typically with longer durations at lower observing frequencies \citep[e.g.][]{rt91,pbc+18} and in some cases no eclipses are seen at all at high-frequency \citep[e.g.][]{sbl+01}. \citet{t+94} present an in-depth review of a range of possible eclipse mechanisms. The authors used observational data of the eclipses of PSR B1957+20 and PSR B1744$-$24A to construct and critically analyse eclipse models, finding that different mechanisms may be responsible for the eclipses in different systems. In the last two decades many more spider pulsars have been discovered, allowing tighter constraints to be placed on possible eclipse mechanisms \citep[e.g.][]{sbl+01,brr+13,rrb+15,bfb+16}, although significant assumptions about the properties of the medium responsible for the eclipses are still required. Some of the more promising mechanisms require magnetic fields in the eclipse medium \cite[e.g.][]{kmg00}, the presence of which is also suggested by depolarisation of pulses near eclipse \citep{ymc+18} and rapid orbital variability possibly attributed to magnetic cycles in the companion \citep{aft94}. On the contrary, interesting recent work by \citet{llm+19} places doubt on the presence of significant magnetic fields near the eclipse boundaries in the BW PSR B1957+20.\\
Spider pulsars are thought to be descendants of low-mass X-ray binaries (LMXBs) after accretion onto the pulsar has ceased \citep{bv91}, a theory which has since been validated through the discovery and observation of systems which transition back and forth between LMXB-like and RB-like states on short timescales \cite[e.g.][]{sah+14}. However, the fate of such systems is much less clear. \citet{rst89rud} postulated that the ablation will eventually lead to complete destruction of the companion star, contributing to the observed family of isolated MSPs, but mass loss rates inferred from observations and modelling appear to be too low for this to occur \citep{el88,s+96}. Aside from evolutionary studies, spider pulsars offer valuable opportunities to investigate the pulsar wind and characteristics of the companion stars under intense irradiation.\\
PSR J2051$-$0827, the second BW discovered \citep{s+96}, is a 4.5\,ms pulsar in a tight, 2.38\,h orbit with a $0.05\,M_{\odot}$ companion \citep[assuming a pulsar mass $1.8\,M_{\odot}$ as in][]{lvt+11}. The system was intensively studied in the few years after its discovery, with radio observations giving insight into it's eclipse properties \citep{s+96,sbl+01} and optical observations showing irradiation of the companion by the pulsar wind, also allowing its orbital inclination to be estimated to be $\sim40^{\circ}$ \citep{sbb96,svl+99,svb+01}. Although thorough, with the data available these early studies left many properties loosely constrained, with little still known about the eclipse mechanism, mass loss and orbital parameters of the system. However, more recent studies are beginning to shed further light on the phenomena; observations of high-energy emission by \cite{wkh+12} suggest the presence of an intra-binary shock between the pulsar wind and the material ablated from the companion star, and continuous long-term radio timing of the pulsar allowed the orbital variations to be mapped over the last two decades \citep{lvt+11,svf+16}.\\
Inspired by the vast amount of data now available on PSR J2051$-$0827 we present here a comprehensive study of the eclipse phenomena over a decade of observations with nearly 4\,GHz of frequency coverage -- the first such study for a black widow system. Details of the observations used are given in Section~\ref{sec:obs} and in Section~\ref{sec:dm} we present long-term measurements of flux density and dispersion measure (DM) modulation. In Section~\ref{sec: rm} we utilise the linearly and circularly polarised flux to investigate the possible presence of magnetic fields within the binary through measurements of RM, Faraday delay and depolarisation. Section~\ref{sec:mechanisms} gives a critique of the eclipse mechanisms, building on the work of \citet{sbl+01}, while Section~\ref{sec: discuss} provides a general discussion of the system in light of the new data.\\
Pertinent to this work, in Fig.~\ref{fig:geometry} we show the expected projection of the binary system on the sky, approximately to scale assuming parameters inferred in the previous studies mentioned above. This inclined view of the system is key to be aware of when inferring parameters of the medium causing eclipses of the pulsar's radio emission.
\noindent

\section{Observations}\label{sec:obs}
\begin{figure*}
	\includegraphics[width=\textwidth]{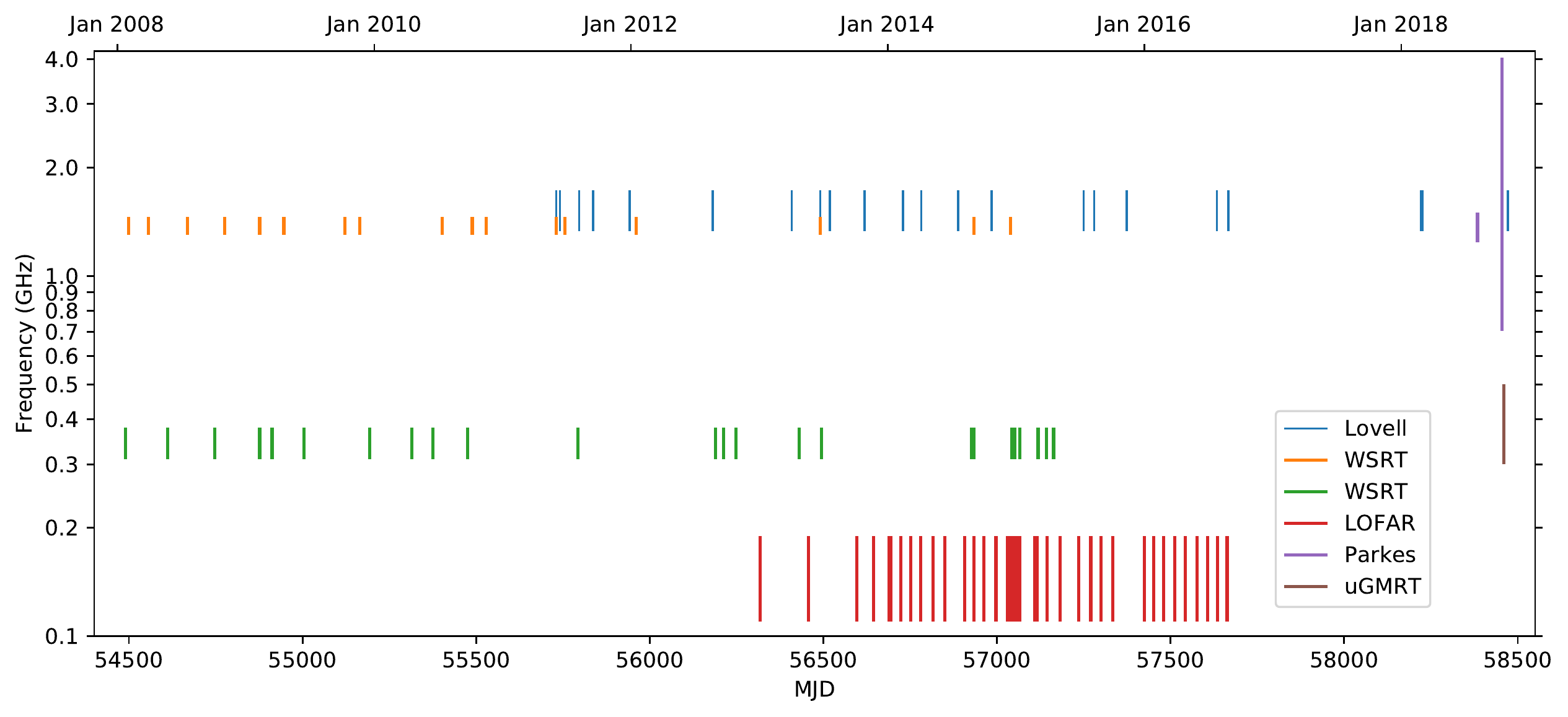}
	\caption[Dates and frequency coverage of observations of PSR J2051$-$0827]{Dates of eclipse observations from each telescope used in this analysis. The $y$-axis location and extent of the markers represent the frequency coverage of the observations. The numerous LOFAR markers correspond to observations at all orbital phases -- not necessarily covering an eclipse -- as the out-of-eclipse data was used for polarisation studies. See Tables~\ref{Table: listobs} and \ref{Table: rms} for exact dates and times.}
	\label{fig:data_dates}
\end{figure*}

The data presented in this paper originate from just over a decades worth of observations of PSR J2051$-$0827 using the Low-Frequency Array \citep[LOFAR;][]{v+13}, Westerbork Synthesis Radio Telescope \citep[WSRT;][]{bh74}, upgraded Giant Metrewave Radio Telescope \citep[uGMRT;][]{gak+17}, Lovell Telescope and Parkes Telescope. The dates and observing frequencies of the specific observations used are given in Fig.~\ref{fig:data_dates} and Tables~\ref{Table: listobs} and \ref{Table: rms}, with observation durations ranging from 15\,min to 3\,h (surpassing the 2.4\,hr orbital period of the binary). The frequency coverage allowed by the telescopes and corresponding backends spans 110--4032\,MHz, with more details given in the following subsections.
\begin{table}
	\centering
	\caption{List of observations of PSR J2051$-$0827 used in this paper. Here the first 5 lines are shown; the full table can be found in the supplementary online material. See Table~\ref{Table: rms} for LOFAR observations.}
	\label{Table: listobs}
	\begin{tabular}{ll}
		\hline
		\multicolumn{2}{c}{WSRT} \\
		\hline
		Date & Frequency \\
		\hline
		16:09 2008-01-25 & 310--380 \\
		08:40 2008-05-25 & 310--380 \\
		20:19 2008-10-08 & 310--380 \\
		12:10 2009-02-14 & 310--380 \\
		08:57 2009-03-22 & 310--380 \\
		\hline
	\end{tabular}
\end{table}\noindent

\subsection{LOFAR}
Observations with LOFAR utilised the high-band antennas (HBA) in $\sim20$--$24$ Core stations. The correlators used to combine the raw complex-voltages from each station were Blue Gene/P and COBALT \citep{bmn+18}, for observations before and after 2014-Apr-18, respectively. The data were recorded in the coherent Stokes mode \citep{s+11}, forming a single tied-array beam from the stations over a 78\,MHz bandwidth, centred at 149\,MHz, with a sampling time of 5.12\,$\upmu$s. These complex-voltage data were run through the LOFAR Known Pulsar Pipeline \citep[PulP; ][]{ahm+10,s+11} using the pulsar ephemeris from the ATNF pulsar catalogue\footnote{http://www.atnf.csiro.au/people/pulsar/psrcat/} \citep{mhth05} to perform coherent dedispersion and fold into \textsc{psrfits}\footnote{https://www.atnf.csiro.au/research/pulsar/psrfits\_definition/Psrfits.html} \citep{hvm04} archive files with sub-integrations of 5\,s duration, 512 pulse phase bins and 400, 195.3\,kHz wide, frequency channels. The resulting folded data were cleaned of RFI using the automated \textsc{paz} method, with `-r' flag, of \textsc{psrchive}\footnote{https://psrchive.sourceforge.net/} \citep{hvm04,vdo12}.\\
During post-processing the folded data were polarisation calibrated using the same technique as described in \citet{nsk+15}. Briefly, this consisted of calculating the Jones matrices for each frequency channel and beam pointing assuming the Hamaker-Carozzi beam model \citep[][\textsc{mscorpol}\footnote{https://github.com/2baOrNot2ba/mscorpol}]{h06} for the HBAs in order to approximate the instrumental response, then using \textsc{psrchive}'s \textsc{pac} and \textsc{pam} methods to apply the calibration solutions and convert the data to Stokes $I, Q, U, V$ parameters, respectively.

\subsection{WSRT}
Each observation made with WSRT was taken in one of two bands; a low-band centred at 345\,MHz with a bandwidth of 70\,MHz, and a high-band centred at 1380\,MHz with a bandwidth of 160\,MHz. The signals from individual antennas were coherently combined in the telescope hardware before being passed to the pulsar backend. All observations utilised the PuMa-\RN{2} backend \citep{ksv08} which recorded the total observed bandwidth in 8 sub-bands, which were later combined for analysis. PuMa-\RN{2} was also used to coherently dedisperse and fold the data with the pulsar ephemeris using \textsc{dspsr}\footnote{http://dspsr.sourceforge.net/} \citep{vb11}, and write to \textsc{psrfits} archive files. The low-band data were folded into archives with 10\,s sub-integrations, 256 pulse phase bins and 448 frequency channels of 156\,kHz width. Conversely, the high-band data were folded into archives with 10\,s sub-integrations, 512 pulse phase bins and 512 frequency channels of 312\,kHz width. Narrowband RFI was identified using a median filter and excised by zero-weighting the contaminated channels using \textsc{psrchive}, and were averaged to 1\,min sub-integrations to reduce their size prior to analysis.

\subsection{Lovell Telescope}
Observations made with the 76\,m, single-dish Lovell Telescope utilised the ROACH backend \citep{bjk+16}. The data were recorded over a 400\,MHz bandwidth centred at 1532\,MHz and were cleaned of RFI in real time using \textsc{dspsr} with the `-skz' flag. Folding of the data used \textsc{dspsr} with the catalogued pulsar ephemeris, and these were written to \textsc{psrfits} archive files with 10\,s sub-integrations, 1600 frequency channels and 512 pulse phase bins. As for the WSRT observations, the archive files were further cleaned of RFI using a median filter.

\subsection{Parkes Telescope}
Two Parkes Telescope observations were obtained, the first covering 3.6\,hrs over 1241--1497\,MHz, using the L-band Multibeam receiver with the Digital Filter Bank Mark \RN{4} (DFB4) back-end, and the second covering 2.7\,hrs over 705--4031\,MHz, using the newly-operational Ultra-Wideband Low-Frequency receiver (UWL) with the Medusa back-end (Hobbs et al., in prep.). The DFB4 back-end recorded data in $8.81$\,$\upmu$s samples which were later split into 1024 channels, folded and written to \textsc{psrfits} files with 512 pulse phase bins and 10\,s sub-integrations. Similarly, the Medusa back-end recorded data in $8.81$\,$\upmu$s samples and were split into 3328 channels, folded and also written to \textsc{psrfits} files with the same pulse phase binning and sub-integration duration as above. For both options the data were provided in four uncalibrated polarisation parameters -- $XX,\ YY,\ \Re[XY],\ \Im[XY]$, which were converted to Stokes parameters using \textsc{pam} in post-processing. \textsc{psrchive}'s \textsc{paz} tool was used for automatic RFI excision.\\ Finally, the data were polarisation calibrated by making use of specific calibration observations made immediately pre- and post-target observation. The calibration observations consisted of pointing the telescope in a direction slightly offset from the target pulsar -- so as to achieve similar background signal while avoiding any bias from the target itself -- and injecting a controlled, periodic noise diode signal at the receiver. The diode signal, polarised at $45^{\circ}$ to the receiver dipoles, could be folded at the known periodicity, mimicking a pulsar. The final measured Stokes parameters could be compared to the known parameters of the injected signal to approximate the Jones matrix \citep{mhb+13}. The Jones matrix solution was then applied to the observation of the target pulsar using \textsc{psrchive}'s \textsc{pac} command.

\subsection{uGMRT}
A single 2\,hr observation with uGMRT was made with all available antennas recording total intensity data over a frequency range of 300--500\,MHz. The raw data from the observations, recorded with a sampling time of $81.92$\,$\upmu$s and frequency resolution of 48.8\,kHz, were converted to filterbank format using the \textsc{filterbank} tool from the \textsc{sigproc}\footnote{http://sigproc.sourceforge.net/} software package. These were then folded and incoherently dedispersed using \textsc{dspsr} with the most recent ephemeris for the pulsar, resulting in data with 4096 frequency channels, 512 pulse phase bins and sub-integrations of 10\,s duration. \textsc{psrchive}'s \textsc{paz} tool was used for automatic RFI excision.

\subsection{Post-processing}
The folded \textsc{psrfits} files for all observations were manually inspected and zapped of any remaining stand-out RFI using the interactive \textsc{psrzap} tool of \textsc{psrchive}. Due to the variable nature of the orbital parameters for PSR J2051$-$0827, all data were modified using \textsc{pam} to install a new ephemeris that corrected for the orbital motion of the pulsar at the time of each observation. For observations made prior to 2014-May the continuous BTX timing model from \citet{svf+16} was used for this. For those data taken after 2014-May the \textsc{tempo2}\footnote{https://sourceforge.net/projects/tempo2/} \citep{hem06} package was used to find pulse time-of-arrivals (TOAs) in 1\,min integrations of the observations, then we fit new timing solutions to TOAs grouped into 6\,month intervals. To perform the fits we initially used the best-fitting T2 model of \cite{svf+16} as a baseline ephemeris, and adjusted the $T0$ and $A1$ parameters -- epoch of periastron and projected semi-major axis of the orbit, respectively -- in \textsc{tempo2} to minimise the RMS residuals of the TOAs for the first 6-month grouping after 2014-May. The timing solution from each 6-month interval was then used as a baseline for the next.\\
Similarly to solving for the variable orbital parameters, we made use of the dispersion measure (DM) values given in \citet{svf+16} to correct for the long-term drift of the out-of-eclipse DMs. For observations made after 2014-May \textsc{psrchive}'s \textsc{pdmp} tool was used to find the DM that maximised the signal-to-noise of the integrated pulsed flux for out-of-eclipse observations in 1\,year intervals.
\noindent

\section{Flux density and dispersion measure}\label{sec:dm}
\citet{s+96, sbl+01} report flux density variations and increases in DM near inferior conjunction of the companion for PSR J2051$-$0827. The flux at low radio frequencies ($< 1$\,GHz) was shown to drop below detection thresholds (i.e. eclipse) regularly at these orbital phases, offering some insight into the physical properties of the material ablated from the companion. With the wealth of data now afforded to us, here we present a much more in-depth investigation of these effects, and their short- and long-term time dependency.\\
\begin{table}
	\centering
	\caption{Step sizes in DM and $\tau$ used in the model fits to data observed at different centre frequencies. $\Delta\tau$ is expressed in units of the pulse period, $P=4.51$\,ms.}
	\label{Table: dm_scat}
	\begin{tabular}{ccc}
		\hline	
		Centre frequency (MHz) & $\Delta$DM (pc\,cm$^{-3}$) & $\Delta\tau$ \\
		\hline
		149 & $2\times10^{-4}$ & $0.005P$ \\
		345 & $4\times10^{-4}$ & $0.01P$ \\
		1380 & $2\times10^{-3}$ & $0.005P$ \\
		1530 & $2\times10^{-3}$ & $0.005P$ \\
		\hline
	\end{tabular}
\end{table}\noindent
To measure the flux density and deviation of DM and scattering timescale, $\tau$, from the out-of-eclipse values, as a function of time, in these observations we employed the same template fitting method as explained in Section 3.1 of \citet{pbc+18}. Briefly, the method consists of creating a high signal-to-noise, 2-dimensional (frequency channel vs. pulse phase bin) template for the data from each telescope by summing multiple out-of-eclipse observations along the time axis, then smoothing out remaining noise with a Savitzky-Golay filter \citep{sg64}. An array of further templates is added by artificially dispersing and scattering the out-of-eclipse template in user-defined steps of $\Delta$DM and $\Delta\tau$, given in Table~\ref{Table: dm_scat} for this work. Least-squares fits of the templates are performed for each sub-integration of data, with the minimum $\chi^2$ taken as the metric to determine the best-fit $\Delta$DM and $\Delta\tau$.\\
Prior to creating the templates and performing the fits, the data were `baselined' by normalising the pulse profile in every frequency channel and time integration by the measured off-pulse noise level. This normalisation means that the scale factor parameter of the best-fit template is directly proportional to the flux density of the pulsar. For observations $> 1$\,GHz we included 4 free parameters across the bandwidth to allow for different scale factors of the sub-bands in order to model the effects of diffractive scintillation -- expected to have a decorrelation bandwidth of $\sim5$--2500\,MHz at frequencies of 1--4\,GHz.

\subsection{L-band}
\begin{figure}
	\includegraphics[width=\columnwidth]{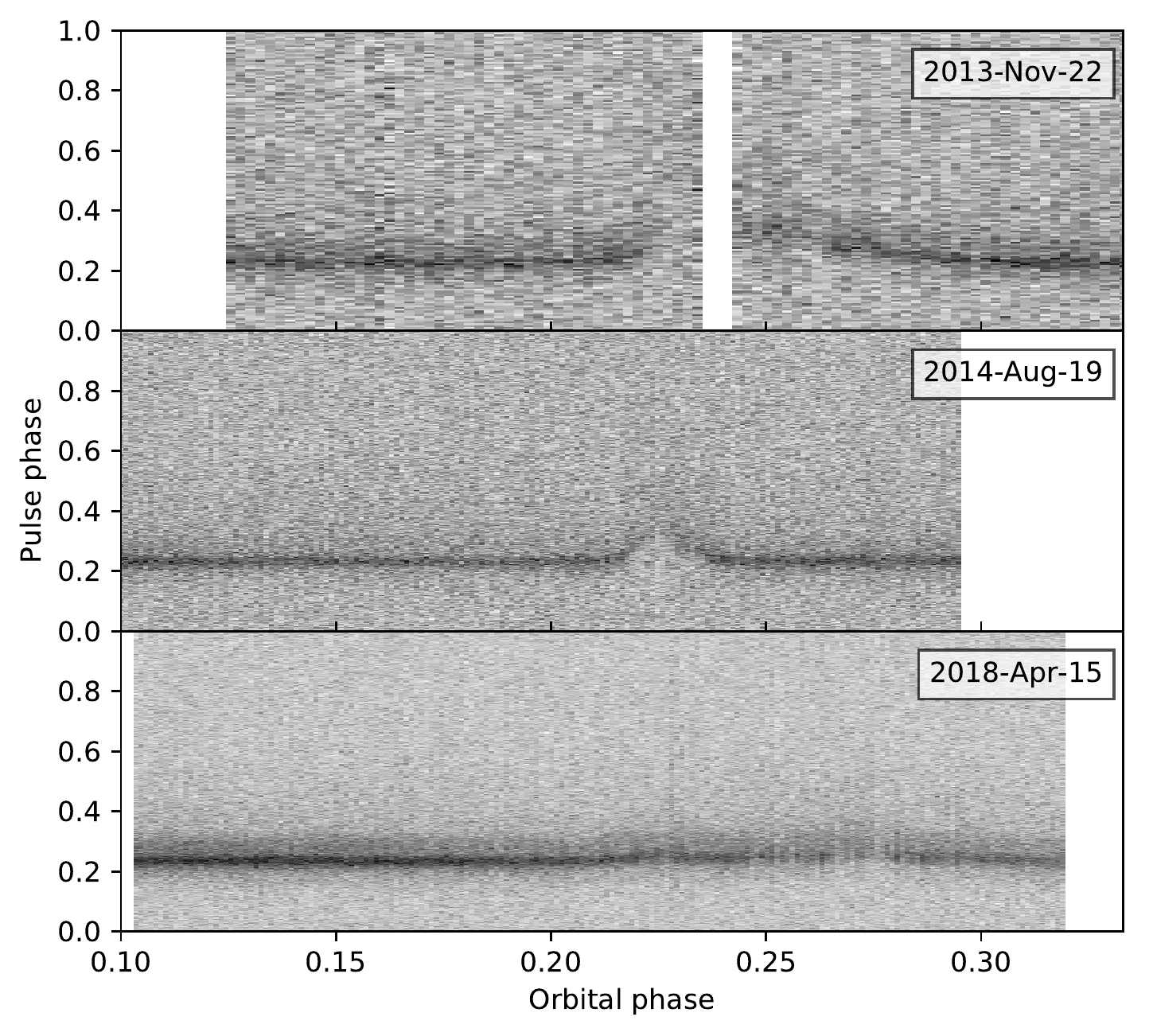}
	\caption[Pulse tracks of PSR J2051$-$0827 for three eclipse traverses]{1530\,MHz observations over the low-frequency eclipse orbital phase range. The greyscale is proportional to intensity, with darker shades corresponding to higher intensities. Delays in the pulse time-of-arrivals can be seen, accompanied by occasional short duration reductions in flux density. Note the distinctly different delay patterns for each of the three dates.}
	\label{fig:Lband_pulse_tracks}
\end{figure}

\begin{figure*}
	\includegraphics[width=\textwidth]{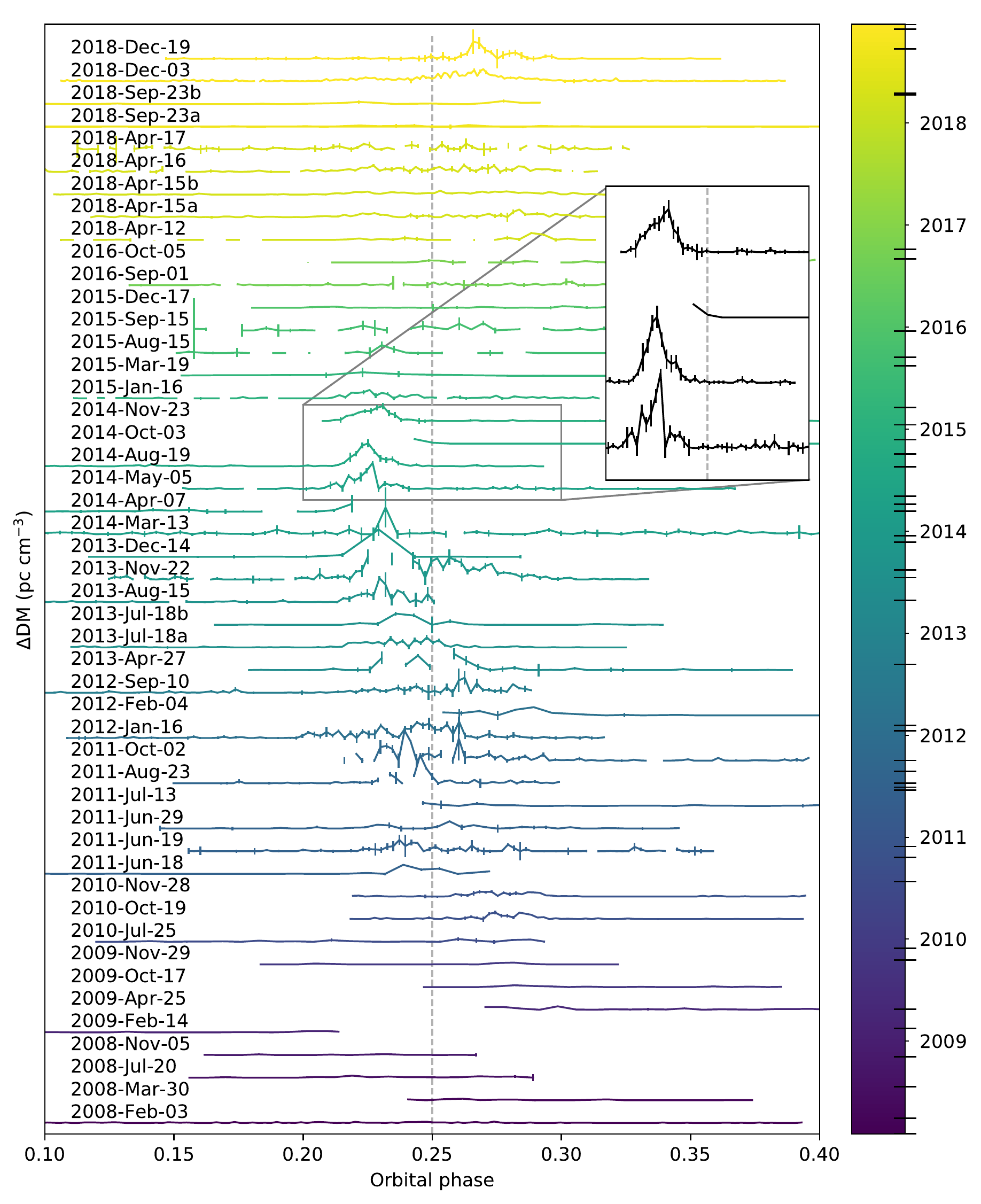}
	\caption[DM variations for all observed eclipse traverses at L-band]{Deviation from mean out-of-eclipse dispersion measures for all observations above 1\,GHz that cover the low-frequency eclipse region. The DMs are all plotted on the same scale, with an offset of $0.2$\,pc\,cm$^{-3}$ applied between consecutive observations. Error bars represent $1\sigma$ uncertainties from the simultaneous DM and scattering fits, as explained in the main text. The dashed grey line marks orbital phase 0.25, where the companion passes closest to our line of sight to the pulsar. The colour of the lines represents the date of observation, with large ticks on the colourbar marking each individual observation on a linear time scale. \textit{Inset}: Zoom in to a 6\,month time period of similar eclipse DM patterns.}
	\label{fig:Lband_dms}
\end{figure*}

With the high frequency WSRT, Lovell Telescope and Parkes Telescope observations covering relatively similar radio frequencies, we treat these as being equivalent for these DM studies. Three L-band observations covering inferior conjunction of the companion, separated by months--years, are shown in Fig.~\ref{fig:Lband_pulse_tracks}. Pulse time-of-arrival (TOA) delays are clearly visible in all three, however each shows distinctly different TOA delay structure. In agreement with \citet{sbl+01} the L-band radio emission is generally detected throughout the orbit, with only sporadic, short duration dips in flux density coincident with the TOA delays in some observations.\\
\noindent
Using the DM and scattering timescale template method detailed above, fits were performed to all L-band observations covering inferior conjunction of the companion. Depending on the signal-to-noise of individual observations the fits were performed on sub-integrations of durations between 20\,s -- 2\,min. The resulting best-fit DMs, along with the corresponding $1\sigma$ uncertainties, are shown in Fig.~\ref{fig:Lband_dms}. The standout feature from these observations is the long-term time evolution of the DM structure in the typical orbital phases corresponding to eclipses. Over the years 2011--2014 we detect significantly more material crossing the line of sight towards the pulsar than either before or after these times, suggesting that trends in the material outflow from the companion can occur on timescales of a few years. The most recent, late-2018, observations appear to show the DMs once again becoming more pronounced, and notably \citet{s+96} show similar pronounced DM profiles in observations made in 1995, suggesting that the period of enhanced DMs seen here is not an isolated `event'. In further detail, specific structure in the DM profiles persists on timescales of months, such as the sharp DM `spike' over the phase range $\sim0.21$--0.25 that regularly occurs in the observations throughout 2014, shown in the inset plot of Fig.~\ref{fig:Lband_dms}. This structure is of specific interest as it occurs prior to inferior conjunction of the companion, leading the star in its orbit. Additionally, the relatively low density of the material means that it would presumably be carried away by the supersonic pulsar wind on timescales much shorter than the orbital period, unless magnetic fields were present to balance the pulsar wind pressure \citep{sbl+01}. These observations support the idea of the presence of magnetic fields throughout the material ablated from the companion, and also suggest that particular magnetic structures can persist in the system for many months. It may be that such variations in the DM profiles are linked to variations in an intra-binary shock, or activity in the companion such as those suggested by \citet{chb18} and \citet{ylk+18}.
\noindent

\subsection{345\,MHz}
\begin{figure*}
	\includegraphics[width=.75\textwidth]{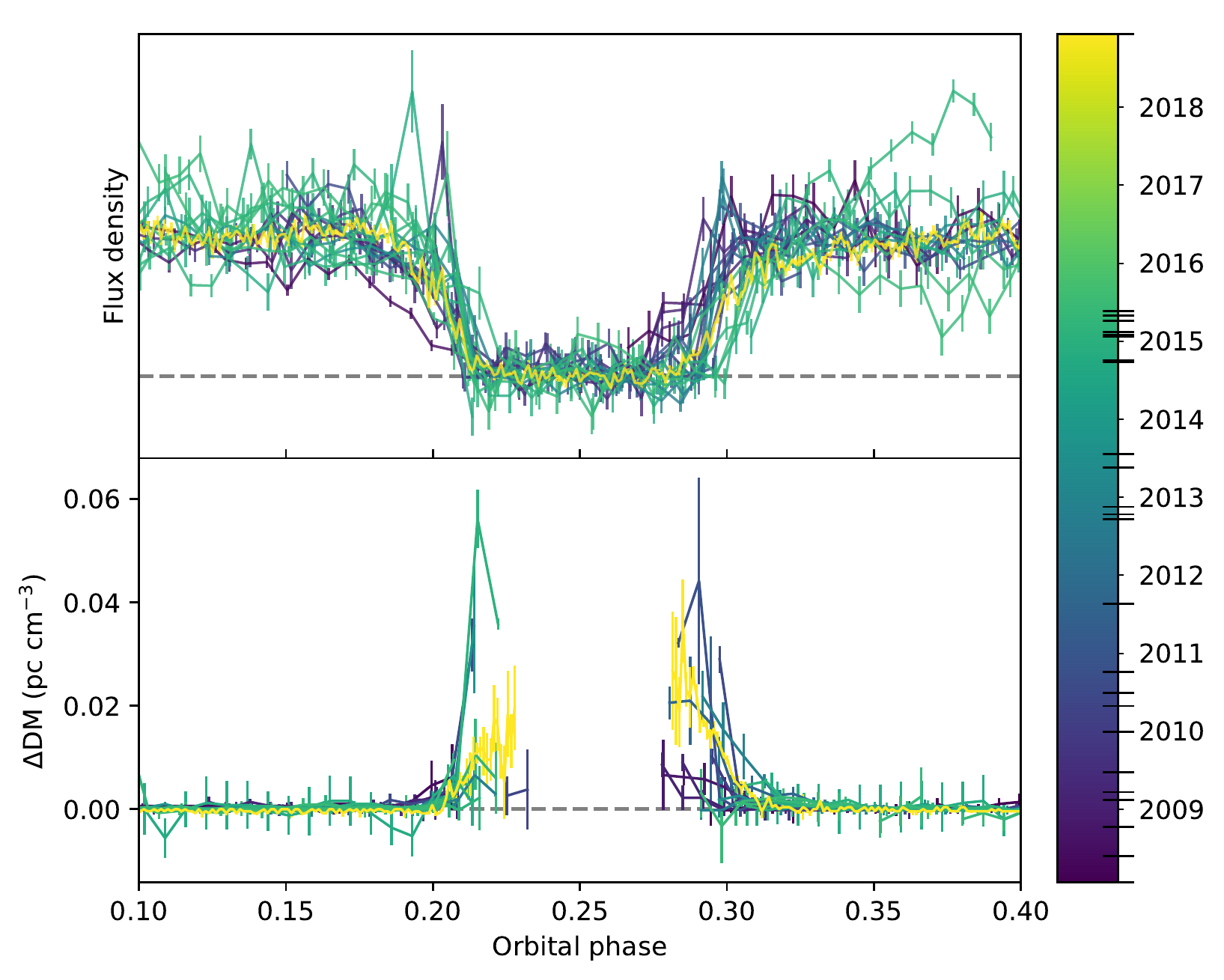}
	\caption[DM and flux density variations for all observed eclipses at 345\,MHz]{\textit{Top}: Measured flux densities for all 345\,MHz observations covering the eclipse region, with each normalised so the the out-of-eclipse mean flux density is unity. The horizontal dashed line corresponds to the detection limit of the telescope. \textit{Bottom}: Deviation from mean out-of-eclipse dispersion measures for the same set of observations. Error bars represent $1\sigma$ uncertainties from the simultaneous DM and scattering fits, as explained in the main text.}
	\label{fig:345_flux_dm}
\end{figure*}

Similar template fits to the WSRT 345\,MHz observations were performed for sub-integrations of 1\,min duration. The resulting best fit DMs and corresponding flux densities are shown in Fig.~\ref{fig:345_flux_dm}. To make the plot clearer, the flux densities have been normalised so that the out-of-eclipse levels are equal to unity, thus removing long-term flux density variations associated with refractive interstellar scintillation. The flux density of the pulsar regularly drops below detection thresholds throughout the companion inferior conjunction, and is never detected between phases $\sim0.23$--0.27, in agreement with the earlier observations of \citet{sbl+01}. There is variability in the shape of the eclipse edges on timescales shorter than 2\,days -- the shortest time interval between observations. Additionally, the large time-span covered by these observations reveals a longer-term trend in the eclipse duration; over the $\sim7.5$\,yr duration the orbital phase of the eclipse egress can be seen to significantly shift, with the eclipses in 2015 persisting to later orbital phases than those previously. There is also some indication of the pulsar re-emerging from eclipse slightly earlier in the 2008 observations, although this is not conclusive. Any variations present in the eclipse ingress are much less prominent than those post-eclipse, being much more stable over time. The extension of the eclipse egress in 2015 relative to the rest of the observations does not have any clear correlation with the higher frequency DM trends over the same date range, shown in Fig.~\ref{fig:Lband_dms}. Note that these eclipse phase shifts are many orders of magnitude larger than the orbital period variations reported in \citet{svf+16}.\\
The deviations in DM in the 345\,MHz eclipse observations, shown in the lower panel of Fig.~\ref{fig:345_flux_dm}, are also variable on timescales shorter than the 2\,day observation interval. In some observations there are sharp rises in DM at the eclipse boundaries, while in others the pulsar falls into, or re-emerges from, eclipse without any significant change in the DM. The deviations in DM generally persist for longer post-eclipse than they do in the ingress, often taking $\sim6$--7\% of the orbit to return to the out-of-eclipse level.\\
The eclipses are generally centred near to phase 0.25, although the extended egresses in 2015 shifts the centre point to a later orbital phase. This shift in the eclipse centre, along with the more prominent flux density variations and slow DM decay in egress are indicative of the ablated material being swept-back due to the orbital motion of the star \citep{fbb+90,sbl+01,pbc+18}.
\noindent

\subsection{149\,MHz}
\begin{figure*}
	\includegraphics[width=.75\textwidth]{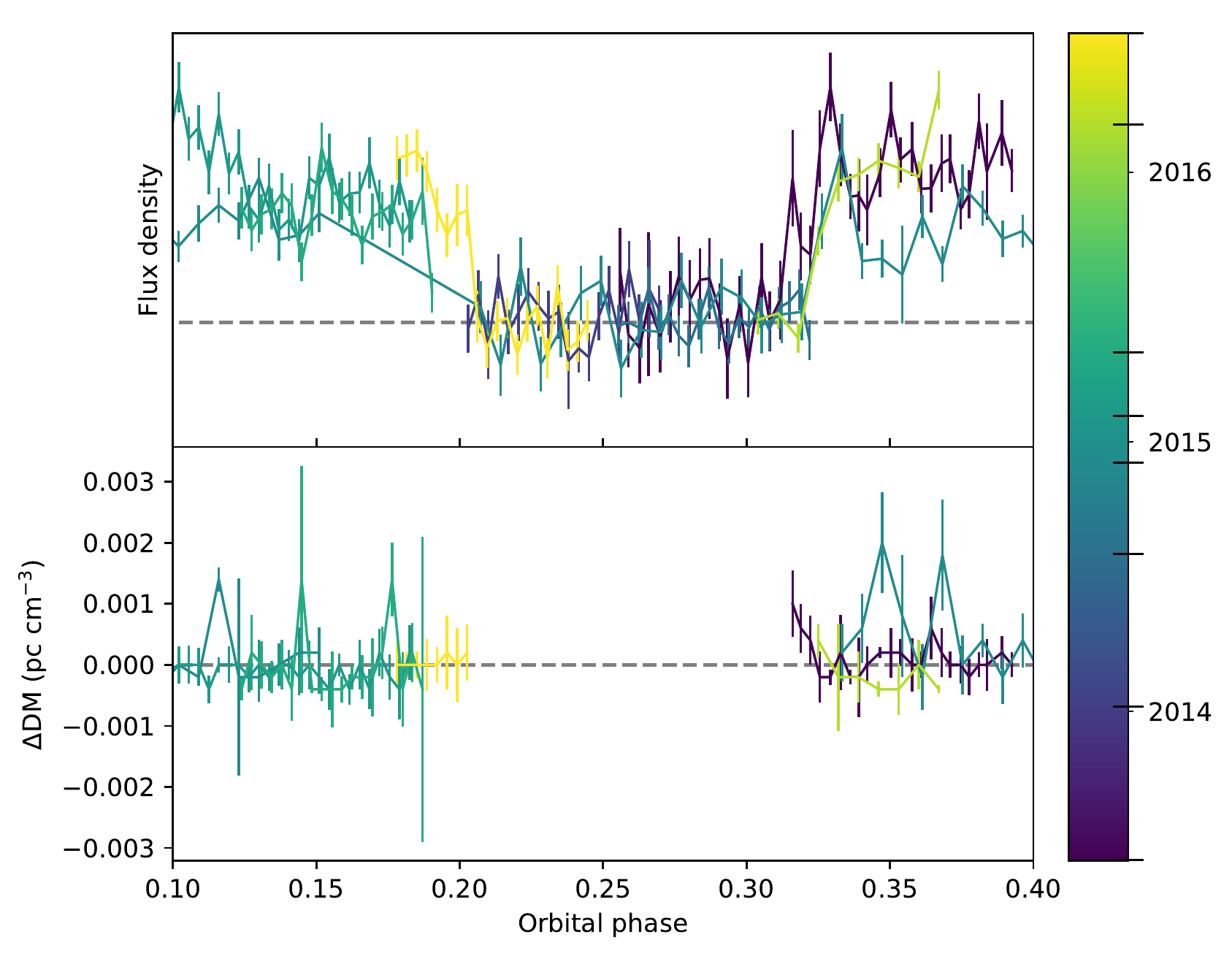}
	\caption[DM and flux density variations for all observed eclipses at 149\,MHz]{\textit{Top}: Measured flux densities for all 149\,MHz observations covering the eclipse region, with each normalised so that the out-of-eclipse mean flux density is unity. The horizontal dashed line corresponds to the detection limit of the telescope. \textit{Bottom}: Deviation from mean out-of-eclipse dispersion measures for the same set of observations. Error bars represent $1\sigma$ uncertainties from the simultaneous DM and scattering fits, as explained in the main text.}
	\label{fig:149_flux_dm}
\end{figure*}

Finally, template fits to 149\,MHz LOFAR observations were performed for sub-integrations of durations between 30--60\,s. The normalised flux densities and deviations of DM from the out-of-eclipse level are shown in Fig.~\ref{fig:149_flux_dm}. As for the 345\,MHz data, the flux densities were normalised such that the out-of-eclipse level was equal to unity, thus removing the effects of long-term refractive interstellar scintillation. The LOFAR eclipse observations span only 3\,yrs; much shorter than the observations at higher frequencies. Moreover, only a handful of observations are available that cover eclipse boundaries, with only one capturing eclipse ingress. In the observations available there are no reappearances of pulsed flux during the eclipse, and the egress orbital phase appears to occur slightly earlier in 2013 than it does in the 2014-Dec and 2016-Mar observations; similar to that seen at 345\,MHz. In general the 149\,MHz egress occurs at a later orbital phase than the 345\,MHz egress, however with only a small amount of data at 149\,MHz, and eclipse-to-eclipse variability larger than the average phase difference between the two frequencies, we defer a full study of the frequency dependence of the eclipse duration to a later paper.\\
The 149\,MHz eclipse ingress occurs without any detectable rise in DM (see Section~\ref{sec:dm_smear} for further discussion), whereas the re-emergence of pulsed flux in the 2013-Jun egress is accompanied by a small additional DM of $\sim0.001$\,pc\,cm$^{-3}$. The centre point of the 149\,MHz eclipse appears to be significantly later than phase 0.25, consistent with a swept-back tail of material extending the eclipse egress \citep{sbl+01}.
\noindent

\section{Polarisation study}\label{sec: rm}
Should the eclipse medium be magnetised, we may be able to detect its influence on the pulsar radio emission as it passes through this material. Previous attempts to detect orbit-dependent polarisation variations have been hampered by the lack of linearly polarised flux from many black widow pulsars, although very recently \citet{ymc+18} observed depolarisation of pulses near eclipse in the globular cluster pulsar PSR J1748$-$2446A (PSR B1744$-$24A; hereafter referred to as Ter5A), which they attribute to both, or either one of, magnetic field variations and multipath propagation through a magnetised, turbulent medium resulting in rapid RM fluctuations. \citet{fbb+90} and \citet{llm+19} instead used circular polarisation measurements of normal, lensed and giant pulses in the black widow PSR B1957+20 to strongly disfavour the presence of strong magnetic fields, either parallel or perpendicular to the line of sight, in the eclipse ingress and egress. However the reliance on detection of pulses meant that no constraints could be made on fields closer to the companion, in the region masked by the eclipse. We investigate the polarisation properties of PSR J2051$-$0827 in this paper, and discuss the possible correlation with orbital phase.

\subsection{Results at low-frequency}\label{sec: rm_low}
\begin{figure}
	\includegraphics[width=\columnwidth]{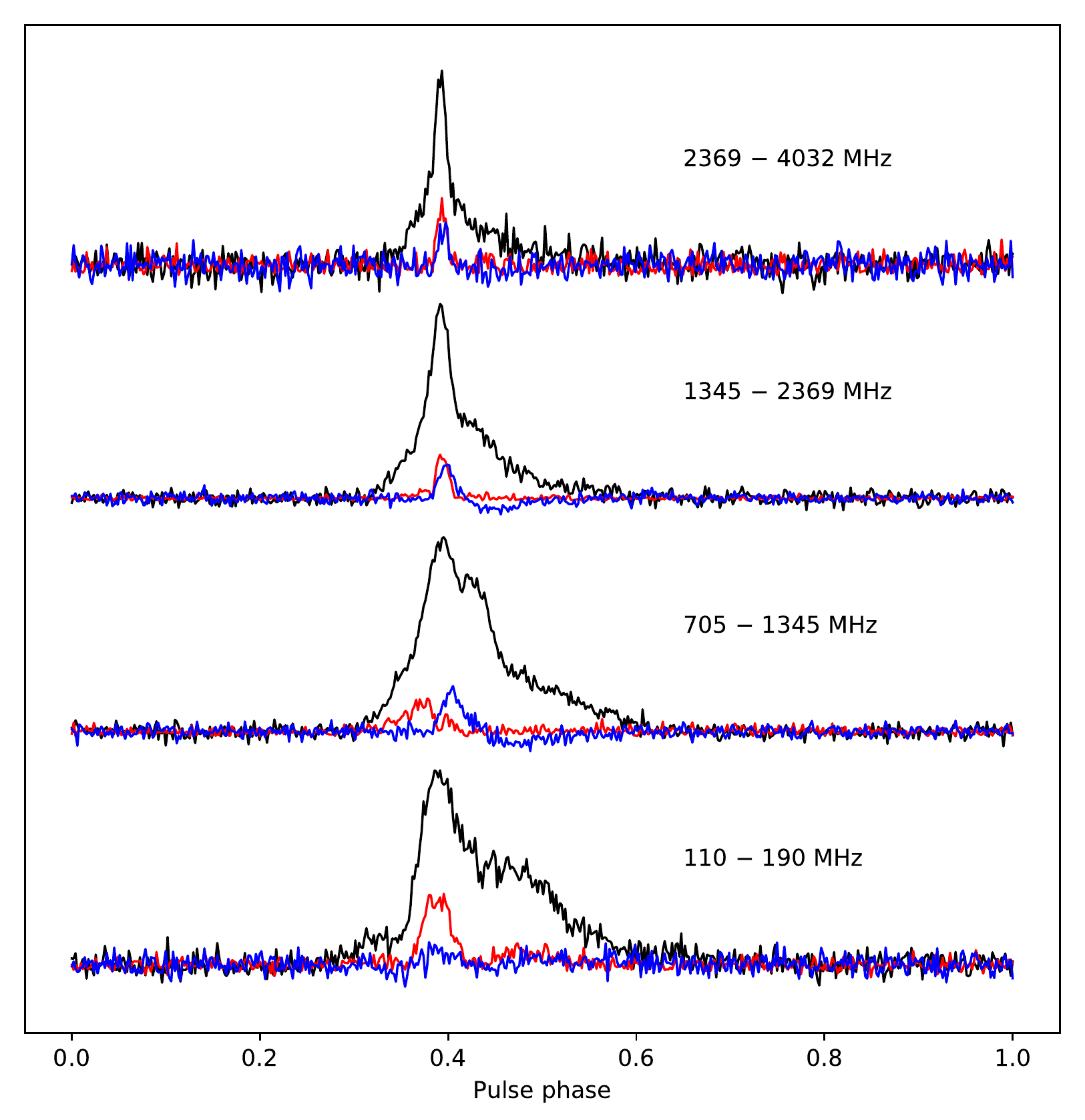}
	\caption[Polarised pulse profiles of PSR J2051$-$0827]{Total intensity, $I$ (black), linear polarisation, $L$ (red), and circular polarisation, $V$ (blue), average pulse profiles of PSR J2051$-$0827 in three simultaneously observed sub-bands using the UWL receiver on the Parkes Telescope, and a low-frequency profile averaged over all out-of-eclipse LOFAR observations. The data were calibrated and RM corrected as explained in the main text.}
	\label{fig:polprofs}
\end{figure}

\begin{figure*}
	\includegraphics[width=\textwidth]{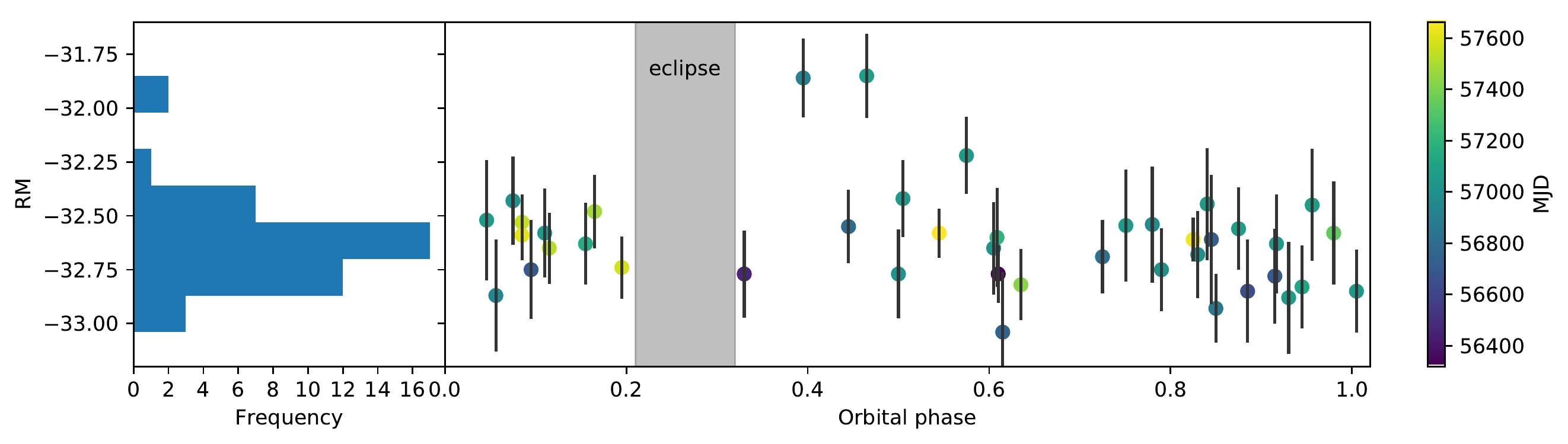}
	\caption[RM measurements as a function of orbital phase for PSR J2051$-$0827]{\textit{Right}: Rotation measure values found to maximise the linear polarisation flux density in 149\,MHz observations, with ionosphere-correction applied (see main text), plotted at the central orbital phase of the corresponding observation. The mean observation duration used to calculate each data point was $\sim 13$\,mins ($0.09P_b$), and the grey region represents the typical eclipse. \textit{Left}: Histogram representing the empirical distribution of the same rotation measure values.}
	\label{fig:rm_meas}
\end{figure*}

The low-frequency, wide-bandwidth nature of LOFAR observations lends itself to studies of the frequency-dependent Faraday rotation of the pulsar radio emission as it traverses any ionised and magnetised medium along the line of sight. This allows for precise measurements of pulsar RM for those pulsars with significant linearly polarised flux densities \citep{sbg+19}.\\
The polarisation calibrated LOFAR data were analysed using \textsc{psrchive}'s \textsc{rmfit} tool. This performs a brute-force search over a user-specified range of RM values, calculating the linearly polarised flux density for each RM and automatically fits a Gaussian to peaks in the resulting spectrum of flux density versus RM. The resulting spectra were manually inspected to check that the fit corresponded to a visually significant peak, and also avoided an occasionally occurring peak at zero RM due to instrumental leakage of the total intensity signal into the orthogonal polarisations. The resulting RM values were ionosphere-corrected using the \textsc{ionFR} code \citep{ssh+13} to calculate the expected ionosphere induced Faraday rotations from total electron content (TEC) ionosphere maps\footnote{ftp://ftp.aiub.unibe.ch/CODE/}, and subtracting these from the measured \textsc{rmfit} values. For observations made prior to 2015 we used the IGRF version 11 geomagnetic field model, and for those observed after this date, beyond which version 11 is no longer valid, we used IGRF version 12\footnote{https://www.ngdc.noaa.gov/IAGA/vmod/igrf.html}. As the \textsc{ionFR} output RMs had a temporal resolution of 1\,hr, and the observation durations ranged from 10--30\,mins, the \textsc{ionFR} values were linearly interpolated to estimate the ionosphere Faraday rotation at the start time of each observation, which was then assumed to be constant for the observation duration. Note that the interpolation always resulted in changes smaller than the $1\sigma$ uncertainties on the \textsc{ionFR} values, thus these approximations do not significantly bias the results. As suggested by \citet{sbg+19}, in Table~\ref{Table: rms} we have included a list of the publicly-available LOFAR data used here, along with the measured RM values and calculated ionosphere corrections, should more precise ionosphere correction methods become available in future.\\
\begin{table*}
	\centering
	\caption{List of LOFAR observations with the corresponding \textsc{rmfit} measured RMs and estimated ionosphere-induced RMs using \textsc{ionFR}. Here the first 5 lines are shown; the full table can be found in the supplementary online material. $^a$No significant RM detection.}
	\label{Table: rms}
	\begin{tabular}{lllcc}
		\hline	
		Date & Project ID & ObsID & Measured RM (rad\,m$^{-2}$) & Ionosphere RM (rad\,m$^{-2}$) \\
		\hline
		11:55 2013-01-26 & LC0\_011 & L85592 & $-28.43\pm0.03$ & $4.34\pm0.13$ \\
		02:54 2013-06-14 & LC0\_011 & L146226 & $-30.87\pm0.11$ & $1.90\pm0.17$ \\
		18:58 2013-10-30 & LC0\_011 & L184310 & N/A$^a$ & $2.36\pm0.22$ \\
		14:34 2013-12-17 & LC1\_027 & L195218 & $-29.79\pm0.06$ & $3.06\pm0.24$ \\
		11:47 2014-02-02 & LC1\_027 & L202628 & $-28.07\pm0.20$ & $4.71\pm0.22$ \\
		\hline
	\end{tabular}
\end{table*}\noindent
Using the ionosphere-corrected RM measurements we calculate PSR J2051$-$0827 to have a weighted mean $\text{RM} = (-32.61\pm0.03)$\,rad\,m$^{-2}$. The RM-corrected linear polarisation profile is shown in Fig.~\ref{fig:polprofs}, and the linear polarisation fraction is found to be $\sim10\%$, which appears to be approximately consistent across observations with high enough signal-to-noise to detect an RM. The calculated RMs for the individual observations are plotted in Fig.~\ref{fig:rm_meas} as a function of orbital phase. The histogram shows that the bulk of the RMs are approximately normally distributed about the mean, and the colourmap shows that there is no clear long-term evolution of the RM. Although we detect no regular rise, or fall, in RM near eclipse, we note that this would not necessarily be expected given the aforementioned highly variable nature of the eclipse material. Instead we highlight a possible increase in variance of RMs post-eclipse, in the orbital phase range $\sim0.3$--0.65. Performing an $F$-test between the egress and all other RMs, with the null hypothesis of equal variances, results in a $p$-value of $5\times10^{-5}$, however this relies on both distributions being near-Gaussian, which is not obviously the case in the egress region. Thus, we perform both a Levene's test \citep{l60} and a Fligner-Killeen test \citep{fk76}, which are more robust to non-normality of the distributions of data \citep{cjj81}. These result in $p$-values of 0.010 and 0.057, respectively. Thus, for the most conservative estimate given by the Fligner-Killeen test, we could expect to observe such data 1 in 20 hypothetical repeats of the observations, assuming that the true variance is the same in both regions. As such this does not strongly favour the case of differing variances in the two regions. However, the lower $p$-values from the $F$ and Levene's tests make it intriguing to investigate this further as more LOFAR data is collected.\\
\noindent
Moreover, assuming that the two highest (least negative) RMs at orbital phases $\sim0.4$ and 0.45 represent physical changes in the Faraday rotation, as opposed to statistical fluctuations, we can use these to estimate the line of sight magnetic field strength in the eclipse medium. Using,
\begin{equation}\label{eq:rm_1}
\text{RM} = \frac{\langle B_{||}\rangle}{1.23 \upmu \text{G}} \text{DM},
\end{equation}
with PSR J2051$-$0827 parameters: out-of-eclipse $\text{DM} = 20.729\text{ pc cm}^{-3}$ and the previously calculated weighted mean $\text{RM} = (-32.61\pm0.03)$\,rad\,m$^{-2}$, we find the mean line of sight magnetic field strength towards the pulsar to be $\langle B_{||}\rangle = (-1.935\pm0.001)$\,$\upmu$G for the out-of-eclipse orbital phases. We attribute this component to the ISM, $\langle B_{||, \text{ISM}}\rangle$, and note that it is comparable to previous measurements of the average ISM field strength \citep[e.g. Fig.~6 of][]{sbg+19}. Then, taking the maximum measured egress $\text{RM} = (-31.85\pm0.19)$\,rad\,m$^{-2}$, Equation~\ref{eq:rm_1} gives the mean line of sight magnetic field strength towards the pulsar to be $\langle B_{||, \text{egress}}\rangle = (-1.890\pm0.008)$\,$\upmu$G, during egress. By assuming that this discrepancy was caused by the medium within the binary system we can break the average magnetic field, DM and RM into components: one corresponding to the medium within the binary, with extent equal to the orbital separation, $a$, and a second corresponding to the ISM, with an extent equal to the distance to the pulsar, $D$ minus the orbital separation:
\begin{eqnarray}
\text{DM} = \int_{0}^{D-a} n_{e, \text{ISM}} \text{d}l + \int_{D-a}^{D} n_{e, \text{sys}} \text{d}l,\\
\text{RM} = \int_{0}^{D-a} n_{e, \text{ISM}} B_{\text{ISM}} \text{d}l + \int_{D-a}^{D} n_{e, \text{sys}} B_{\text{sys}} \text{d}l,
\end{eqnarray}
where the subscripts ISM and sys correspond to the components along the line of sight towards the binary and within the binary, respectively. By approximating the medium within each region to be homogeneous, these reduce to:
\begin{eqnarray}
\text{DM} \approx (D-a)\langle n_{e, \text{ISM}}\rangle + a\langle n_{e,\text{sys}}\rangle,\\
\text{RM} \approx (D-a)\langle n_{e, \text{ISM}}\rangle\langle B_{||,\text{ISM}}\rangle + a\langle n_{e, \text{sys}}\rangle\langle B_{||, \text{sys}}\rangle.
\end{eqnarray}
Thus, using Equation~\ref{eq:rm_1}, the average line of sight magnetic field strength over the full distance to the pulsar is given by
\begin{equation}\label{eq:rm_2}
\langle B_{||}\rangle = \frac{(D-a)\langle n_{e, \text{ISM}}\rangle\langle B_{||, \text{ISM}}\rangle + a\langle n_{e, \text{sys}}\rangle\langle B_{||, \text{sys}}\rangle}{(D-a)\langle n_{e, \text{ISM}}\rangle + a\langle n_{e, \text{sys}}\rangle}.
\end{equation}
There are two scenarios to consider: one with additional electron density in the system at egress orbital phases, and one without. These can give estimations of lower and upper limits, respectively, on the magnetic field within the binary at these orbital phases, assuming the RM deviations are attributed entirely to the intra-binary medium.\\
Firstly, taking the case of no additional material in the system, we have $\langle n_{e, \text{sys}}\rangle = \langle n_{e, \text{ISM}}\rangle$. Equation~\ref{eq:rm_2} then simplifies to
\begin{equation}
\langle B_{||}\rangle = \frac{1}{D}\left( (D-a)\langle B_{||, \text{ISM}}\rangle + a\langle B_{||, \text{sys}}\rangle\right),
\end{equation}
thus, by rearranging we find the mean line of sight magnetic field strength within the binary to be
\begin{equation}\label{eq:rm_3}
\langle B_{||, \text{sys}}\rangle = \frac{1}{a}\left( D\langle B_{||}\rangle - (D-a)\langle B_{||, \text{ISM}}\rangle\right).
\end{equation}
Now, if we take the scenario where extra material is present within the binary, we can manipulate Equation~\ref{eq:rm_2} to explicitly show that the system material is made up of an equivalent ISM electron density, plus a term relating to the additional eclipse material, i.e. $\langle n_{e, \text{sys}}\rangle = \langle n_{e, \text{ISM}}\rangle + \langle \Delta n_{e}\rangle$. Equation~\ref{eq:rm_2} then leads to,
\begin{equation}
\langle B_{||}\rangle = \frac{(D-a)\langle n_{e, \text{ISM}}\rangle\langle B_{||, \text{ISM}}\rangle + a\left(\langle n_{e, \text{ISM}}\rangle + \langle\Delta n_{e}\rangle\right)\langle B_{||, \text{sys}}\rangle}{D\langle n_{e, \text{ISM}}\rangle + a\langle\Delta n_{e}\rangle},
\end{equation}
giving,
\begin{equation}\label{eq:rm_4}
\langle B_{||, \text{sys}}\rangle = \frac{\left(D\langle n_{e, \text{ISM}}\rangle + a\langle\Delta n_{e}\rangle\right)\langle B_{||}\rangle - (D-a)\langle n_{e, \text{ISM}}\rangle\langle B_{||, \text{ISM}}\rangle}{a\left(\langle n_{e, \text{ISM}}\rangle + \langle\Delta n_{e}\rangle\right)},
\end{equation}
where we have $\langle n_{e, \text{ISM}}\rangle = \frac{\text{DM}_{\text{out-of-eclipse}}}{D}$ and $\langle\Delta n_{e}\rangle = \frac{\Delta\text{DM}}{a}$, with $\Delta\text{DM}$ given by the LOFAR measurements in Section~\ref{sec:dm}. As the LOFAR observations show no detectable increase in DM at orbital phases 0.4--0.45, we can place an upper limit corresponding to the typical $1\sigma$ uncertainties on the LOFAR $\Delta\text{DM}$ measurements of 0.001\,pc\,cm$^{-3}$. Taking relevant parameter values of $\text{DM}_{\text{out-of-eclipse}} = 20.729\text{ pc cm}^{-3}$, $D = 1.47 \text{kpc}$ \citep[DM-derived distance using the electron density model of][]{ymw17}, $a = 1.1 R_{\odot}$ (from the orbital parameters given in Section~\ref{sec: intro}), $\langle B_{||, \text{ISM}}\rangle = (-1.935\pm0.001)$\,$\upmu$G, $\langle B_{||}\rangle = (-1.890\pm0.008)$\,$\upmu$G and accounting for only the uncertainties on $\langle B_{||, \text{ISM}}\rangle$ and $\langle B_{||}\rangle$ using standard error propagation, Equations~\ref{eq:rm_3} and \ref{eq:rm_4} loosely constrain the mean line of sight magnetic field strength within the binary to be $9(2)\times10^{-4}$\,$\text{G} \lesssim \langle B_{||, \text{sys}}\rangle \lesssim 2.7(5)\times10^3$\,G, where the number in brackets represents the uncertainty on the last digit, should an intra-binary field be responsible for the observed change in RM at orbital phase 0.4.
\noindent

\subsection{Results at high-frequency}
Observing at high-frequencies offers an invaluable complement to the LOFAR observations discussed above; although the measurements are much less sensitive to small variations in RM, the radio pulsations from the pulsar are detectable throughout the entire eclipse region. As such, they provide a probe into the medium directly responsible for the eclipses at low-frequencies. Two observations were recently performed using the Parkes Telescope, with the primary aim to measure the polarisation properties of PSR J2051$-$0827 throughout its orbit. The first spanned 1241--1497\,MHz and covered $\sim1.5$ consecutive eclipses, while the second spanned 705--4032\,MHz and covered 1 eclipse, offering an unprecedented simultaneous wide-band view of the eclipse region. The resulting polarisation calibrated data were RM corrected using a value of $\text{RM} = (-33.1\pm1.0)$\,rad\,m$^{-2}$, measured with \textsc{rmfit} on the out-of-eclipse, wide-band data.\\
The data, in the form of the $I, Q, U, V$ Stokes parameters, were dedispersed in each sub-integration and frequency channel using the DM values measured through the template fitting technique discussed in Section~\ref{sec:dm}, thus removing the effects of increased dispersion during the eclipse. In the absence of any flux calibrator observations, the data were normalised to reduce any instrumental effects on the flux densities. This normalisation consisted of two steps: firstly, the mean off-pulse levels were subtracted from the corresponding profiles in each frequency channel, sub-integration and Stokes parameter, and secondly, the profiles were divided by the standard deviation of the corresponding off-pulse regions of the Stokes $I$, total intensity, data. At this stage, the RM-corrected wide-band observation data were split into 3 sub-bands, and each summed along the frequency axis for analysis. Finally, the Stokes $Q$ and $U$ profiles in each sub-integration were combined in quadrature to give the total linear polarisation profiles as a function of time. The resulting total intensity, $I$, linear, $L$, and circular, $V$, average profiles are shown in Fig.~\ref{fig:polprofs} for the 3 sub-bands.

\subsubsection{Depolarisation}
\begin{figure*}
	\includegraphics[width=\textwidth]{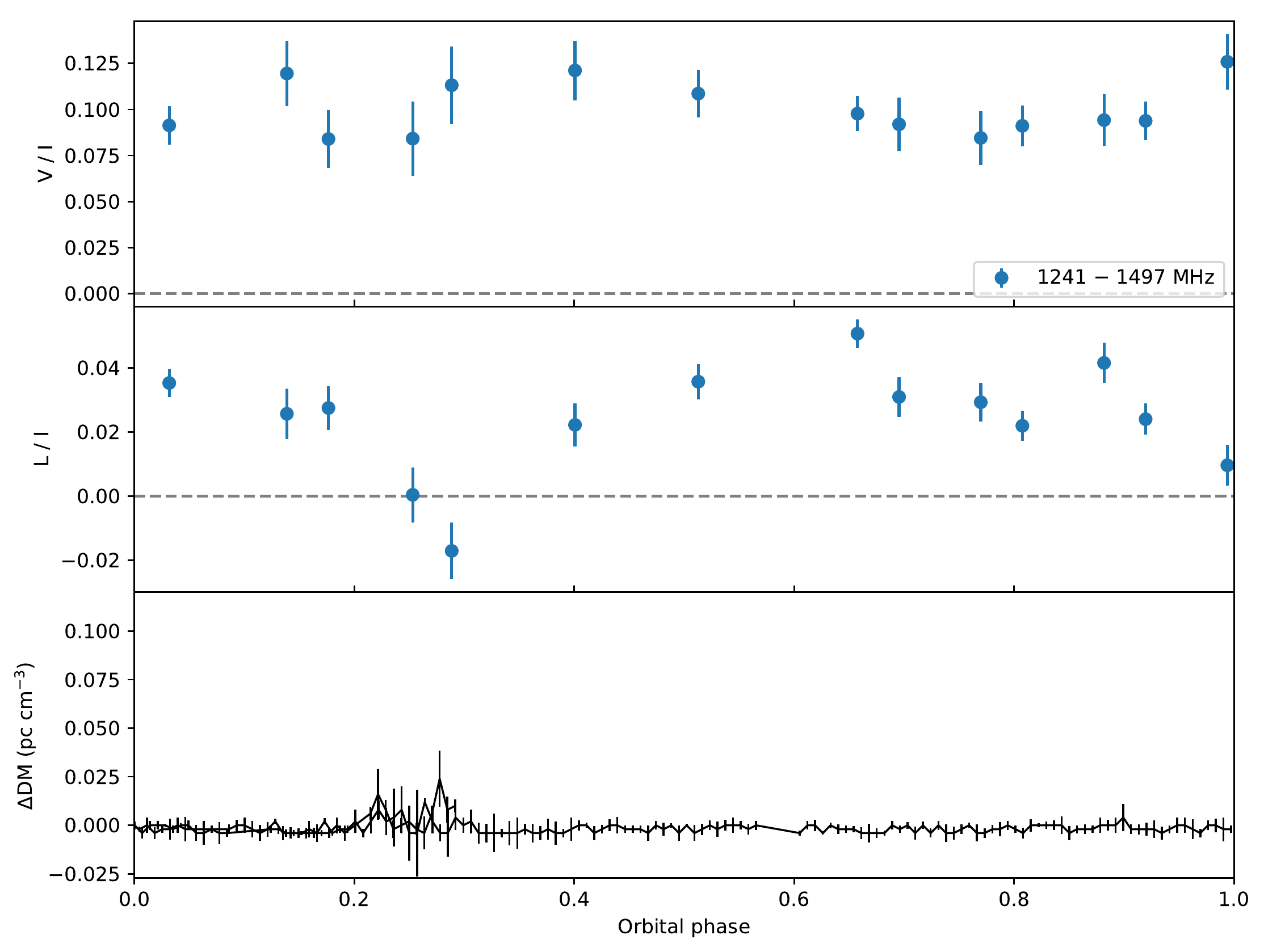}
	\caption[Polarised flux as a function of orbital phase for PSR J2051$-$0827 as observed on 2018-Sep-23]{Circularly (\textit{top}) and linearly (\textit{middle}) polarised fractions of the total intensity flux of PSR J2051$-$0827 measured in a single frequency band, as a function of the pulsar's orbital phase. \textit{Bottom}: Measured DM relative to the out-of-eclipse mean for the same observation. The DM vertical axis scale has been set to match that of Fig.~\ref{fig:polflux_uwl} to highlight the relative $\Delta$DM magnitudes. Observed on 2018-Sep-23 with the Parkes Telescope.}
	\label{fig:polflux_dfb}
\end{figure*}

\begin{figure*}
	\includegraphics[width=\textwidth]{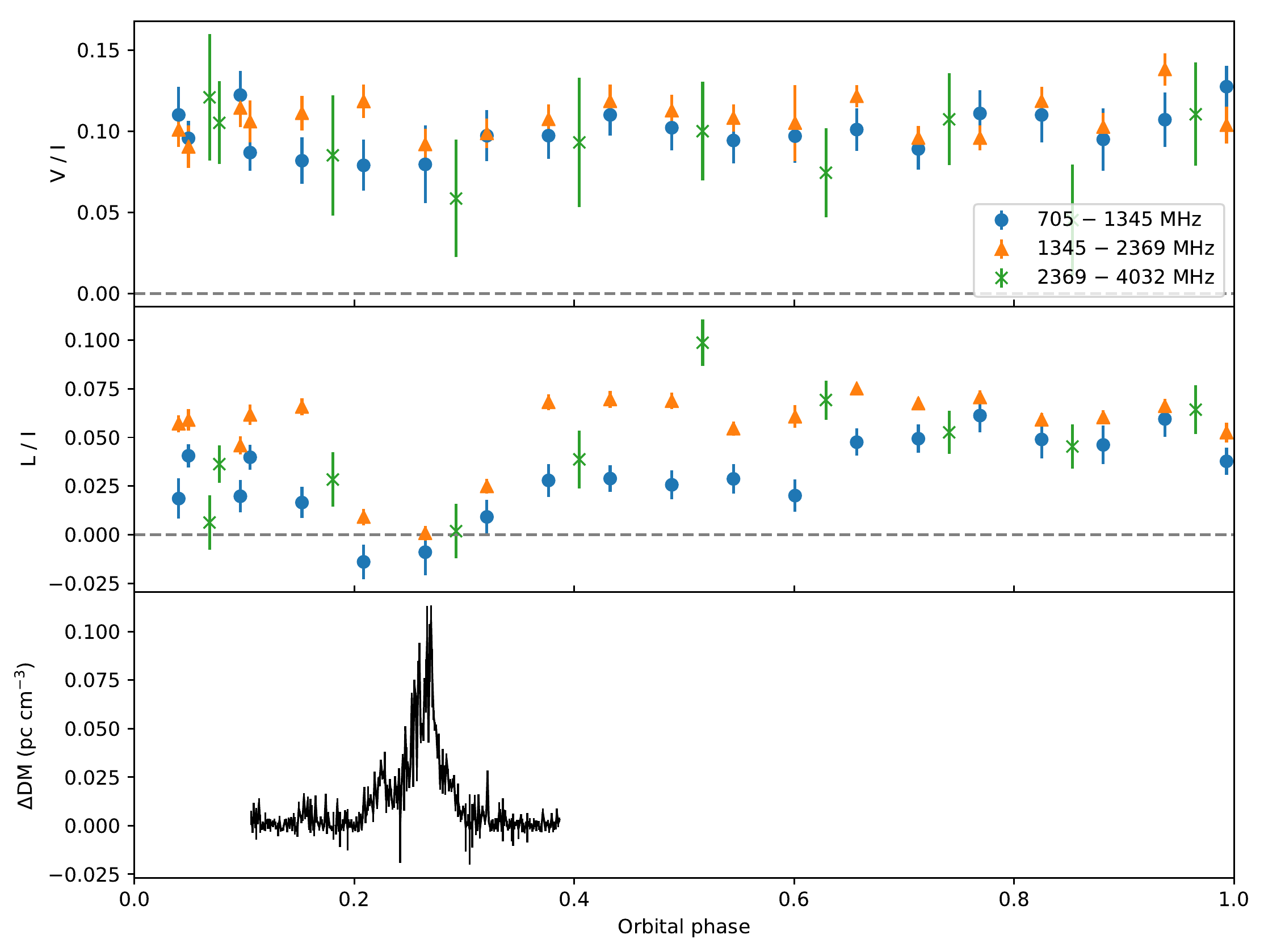}
	\caption[Polarised flux as a function of orbital phase for PSR J2051$-$0827 as observed on 2018-Dec-03]{Circularly (\textit{top}) and linearly (\textit{middle}) polarised fractions of the total intensity flux of PSR J2051$-$0827 measured in 3, simultaneously observed, frequency sub-bands, as a function of the pulsar's orbital phase. \textit{Bottom}: Measured DM relative to the out-of-eclipse mean for the same observation, utilising the entire 705--4032\,MHz bandwidth. Observed on 2018-Dec-03 with the Parkes Telescope.}
	\label{fig:polflux_uwl}
\end{figure*}

\begin{figure*}
	\includegraphics[width=\textwidth]{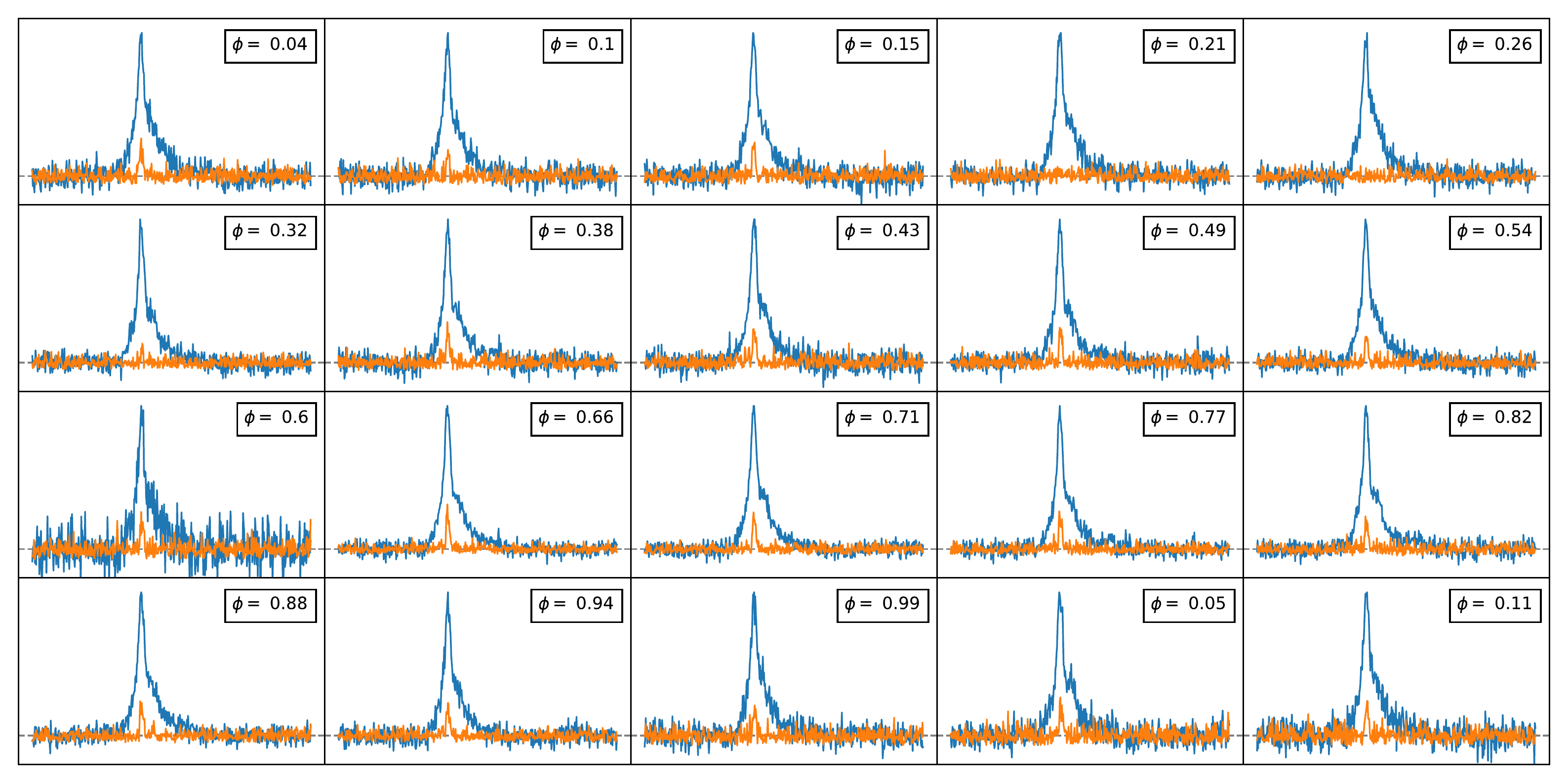}
	\caption[Polarised pulse profiles as a function of orbital phase for PSR J2051$-$0827]{Total intensity (blue) and linear polarisation (orange) pulse profiles in integrations of 8\,mins for a Parkes Telescope observation of PSR J2051$-$0827 on 2018-Dec-03. The profiles are integrated over the sub-band with the highest signal-to-noise, 1345--2369\,MHz. The orbital phase at the centre point of each integration is indicated in the top-right corner of each sub-plot. To aid visualisation the axes are consistent for each sub-plot, with the full pulse profiles (phases 0--1) displayed along the x-axes. The zero level is indicated by a horizontal dashed line and the total intensity and linear polarisation pulse profiles are scaled by a constant for each plot such that the total intensity profile has a maximum value of unity, thus removing the effect of variations in total intensity.}
	\label{fig:depolprofs}
\end{figure*}

In order to test for depolarisation of the pulses near eclipse, such as those reported in \citet{ymc+18} for Ter5A, we measure the flux densities of $L$ and $V$ as a function of the pulsar's orbital phase. The average $I$, $L$ and $V$ profiles were smoothed using a Savitzky-Golay filter (similar to the method in Section~\ref{sec:dm}) to create one-dimensional profile templates. The relative flux density in each sub-integration was determined through a least-squares fit of the template amplitude (multiplication factor). To remove the influence of variations in the total intensity flux, each of the $L$ and $V$ amplitudes were divided by the corresponding amplitude of the $I$ fits. Finally, the relative amplitudes were multiplied by the ratio of the sum over the $L$ (or $V$) template profile to the sum over the $I$ profile in order to give the absolute polarisation fractions. The resulting polarisation fractions as a function of orbital phase are shown in Figs.~\ref{fig:polflux_dfb} and \ref{fig:polflux_uwl} for the first and second observations, respectively, along with the corresponding $\Delta$DM measurements found previously.\\
\noindent
\noindent
In both observations the circular polarisation fraction is shown to persist at a constant level of $\sim10\%$, within uncertainties, throughout the full orbit. In contrast, the linear polarisation fraction reduces to zero during the orbital phase range $0.15 \lesssim \phi \lesssim 0.35$, corresponding to the typical low-frequency eclipse region. It is interesting to note that this is the case in both observations, despite the much smaller DM increases in the $\sim1.5$ eclipses in the first observation. The evidence is also strengthened by the finding that the depolarisation is measured independently in the three wide-spread sub-bands in the second observation. We note that this depolarisation was also discovered independently in Effelsberg data by Oslowski et al. (in prep.). Due to the relatively low signal-to-noise of the narrow linear polarisation profile, it may be possible that scattering could smear out the profile, as the pre-processing of the data only corrected for the variable DMs. However, as shown in Fig.~\ref{fig:depolprofs}, the total intensity profiles at orbital phases $\phi = 0.21, 0.26, 0.32$ show only minimal scatter broadening in comparison to other phases, whereas the linear polarisation profile is not visible at all at phases $\phi = 0.21, 0.26$, and begins to reappear at phase $\phi = 0.32$.\\
\noindent
An alternative possibility is that the RM increases, decreases or fluctuates such that the integrated linear polarisation profile, corrected with the `wrong' RM is smeared out. As an unfortunate consequence of the low signal-to-noise of the linearly polarised flux density, we find that even in the brightest sections of the observations more than 10\,mins of data must be integrated in order to reliably detect a peak in the RM spectrum produced by \textsc{rmfit}. Such long integrations span nearly the full duration of the eclipse region, thus our measurements are restricted to attempting to detect either an increase or decrease in RM, which remains at a nearly constant value throughout the eclipse region, before returning to the out-of-eclipse level. Using $\Delta\phi = \text{RM} \lambda^2$, where $\Delta\phi$ is the polarisation position angle rotation in radians and $\lambda$ the observing wavelength in metres \citep{lk04}, one can calculate the maximum detectable RM in a dataset through the requirement that the polarisation position angle must turn through $< 2\pi$ within a frequency channel in order to avoid complete smearing. In the data used here the lowest frequency channel spans 704--705\,MHz, which gives a maximum $\text{RM} \sim \pm 12000$\,rad\,m$^{-2}$ in order for the intra-channel position angle rotation to remain $< 2\pi$. Using \textsc{rmfit} to search a 16\,min integration over the dedispersed eclipse region, we find no significant peak in the RM spectrum anywhere in this range. In contrast, searches in all other 16\,min integrations throughout the orbit detect significant peaks consistent with the average out-of-eclipse RM value. Using Equation~\ref{eq:rm_4}, assuming values applicable to these observations of $\Delta \text{DM} \approx 0.1$\,pc\,cm$^{-3}$ and an eclipse $\text{RM} \sim 12000$\,rad\,m$^{-2}$, the inferred intra-binary magnetic field would be $\sim10^7$\,G. As any RM larger than this would require even more unrealistically large magnetic fields, the possibility of an increased (or decreased), but constant, RM within the eclipse region can be ruled out.\\
These measurements imply that, as for Ter5A \citep{ymc+18}, the pulsar flux from PSR J2051$-$0827 becomes depolarised as it traverses the intra-binary medium responsible for the low-frequency eclipses. \citet{ymc+18} postulate that the depolarisation occurs as a result of rapid RM fluctuations in a magnetised turbulent medium causing smearing out of the linear polarisation. In light of our observations, and their similarity to those attained for Ter5A, it appears plausible that the same mechanism is responsible here. As such, it is fruitful to follow their analysis using values applicable to these observations.\\
By approximating the inferred RM fluctuations, and thus polarisation position angle fluctuations, to be normally distributed with standard deviations of $\sigma_{\text{RM}}$ and $\sigma_{\psi} = \lambda^2 \sigma_{\text{RM}}$, respectively, \citet{ymc+18} derive the depolarisation resulting from integration over the fluctuations to follow,
\begin{equation}\label{eq: depol}
L = L_0 \exp\left(-2 \lambda^4 \sigma_{\text{RM}}^2 \right),
\end{equation}
where $L_0$ and $L$ represent the linear polarisation magnitude before and after the integration over the fluctuations, respectively, and $\lambda$ is the observing wavelength. Since all three of the observing sub-bands become depolarised for the second observation of PSR J2051$-$0827, the tightest constraint that can be placed is a lower limit on the magnitude of the RM fluctuations that would be necessary to depolarise the highest-frequency radio emission under these assumptions. Taking the depolarised fraction to be equal to the mean $1\sigma$ uncertainties on the high-frequency measurements, $L_0 / I \approx 0.01$, the polarised fraction to be equal to the mean out-of-eclipse value, $L / I \approx 0.045$, and the observing wavelength to be $\lambda \approx 9$\,cm, Equation~\ref{eq: depol} gives the threshold standard deviation of RM fluctuations of $\sigma_{\text{RM}} \gtrsim 100$\,rad\,m$^{-2}$. Further, by assuming that DM fluctuations from the medium are also normally distributed, with standard deviation $\sigma_{\text{DM}} \sim 0.01$\,pc\,cm$^{-3}$, then Equation~\ref{eq:rm_1} can be modified to give
\begin{equation}
\frac{B_{||}}{1.23\upmu\text{G}} = \left( \frac{\sigma_{\text{RM}}}{\sigma_{\text{DM}}} \right)^{1/2}.
\end{equation}
Using values relevant to PSR J2051$-$0827 implies that the mean magnetic field parallel to the line of sight in the eclipse medium, $B_{||} \gtrsim 10^{-4}$\,G. This is around an order of magnitude larger than the lower limit placed on Ter5A through the same method \citep{ymc+18}. Assuming that such a field would be provided by a dipolar magnetic field originating at the companion, and taking the orbit to be inclined at $40^{\circ}$ \citep{svb+01}, with further orbital parameters as in \citet{svf+16} and \citet{lvt+11}, the closest approach of the line of sight to the companion centre is $\approx0.83 R_{\odot}$. With a Roche lobe radius of $R_L \approx 0.15R_{\odot}$, the implied surface magnetic field strength of the companion is $B_{\text{S}}  = B_{||} \left(\frac{0.83}{0.15}\right)^3 \gtrsim 2 \times 10^{-2}$\,G (see also Section~\ref{sec:faraday_delay}).\\
Note also that this rules out the possibility of a completely balanced pair plasma, which could be produced if the eclipse material was predominantly fed by pulsar wind particles \citep{llm+19}. In the case of a pair plasma, the induced rotations of the linear polarisation vector would be equal and opposite for the positive and negative charged particles, and as such, no net rotation would occur and the linear polarisation would not be depolarised.

\subsubsection{Persistent circular polarisation}
Recent work by \citet{llm+19}, in which the authors placed strong upper limits on the presence of magnetic fields near the eclipse region in the black widow pulsar PSR B1957+20, highlighted the usefulness of circular polarisation measurements for investigating the effects of intra-binary magnetic fields. Their study suggested that \textit{generalised Faraday rotation} \citep{km98} may become important in the context of spider pulsars. Specifically, in the case of a strong magnetic field directed near-perpendicular to the line of sight the natural propagation modes through the plasma become strongly elliptical, or entirely linear for a completely perpendicular field. This contrasts with the two circular polarisation natural modes corresponding to a weak, or near-parallel, magnetic field which are relevant to the above investigation of `normal' Faraday rotation.\\
For circular natural modes Faraday rotation refers to the rotation of the linear polarisation vector around the natural axis of the plasma, which has a dependence on the frequency of the radiation and thus if not corrected for will effectively smear out the linear polarisation when integrated over a finite bandwidth, while leaving the circular polarisation unaffected. Alternatively, for elliptical natural modes the differential rotation of the polarisation vector occurs in the linear \textit{and circular} planes, and can act to depolarise both when integrated over a finite bandwidth. Finally, for the case of entirely linear natural modes, i.e. a strong perpendicular field, the rotation occurs in the circular polarisation domain, causing this to depolarise while leaving the linear polarisation unaffected.\\
Considering the case of a strong near-perpendicular field, the differential rotation of the linear `$x$' and `$o$' natural modes can be approximated \citep{t+94,llm+19} to lead to a phase difference between the two modes:
\begin{equation}
\Delta\Phi_{x,o} \approx \Delta\Phi_{\text{DM}}\frac{\left< f_{B_{\perp}}^2 \right>}{f^2},
\end{equation}
where $\left< f_{B_{\perp}}^2 \right>$ represents the electron-density weighted average of the (squared) cyclotron frequency, $f_{B_{\perp}} = \frac{qB_{\perp}}{2\pi m_e} \approx 2.8 \left( \frac{B_{\perp}}{1\,\text{G}}\right)$\,MHz, $f$ is the observation frequency, and $\Delta\Phi_{\text{DM}} = 2\pi k_{\text{DM}} \frac{\Delta\text{DM}}{f}$ is the additional dispersion-induced phase delay, with \newline$k_{\text{DM}} = \frac{e^2}{2\pi m_e c} \approx 4149$\,s\,MHz$^2$\,cm$^3$\,pc$^{-1}$. In the presence of a sufficiently strong field, the phase difference between the two modes will vary over the observing band, causing the circular polarisation vector to rotate and effectively depolarise when integrated over frequency. Thus, the fact that we observe no significant depolarisation of Stokes $V$ with orbital phase in Figs.~\ref{fig:polflux_dfb} and \ref{fig:polflux_uwl} suggests that the differential phase difference between the top and bottom of the band is less than one full rotation, i.e. $\lesssim2\pi$. As such, we can derive:
\begin{eqnarray}
\Delta\Phi_{x,o,\text{low}} - \Delta\Phi_{x,o,\text{high}} \lesssim 2\pi\\
\left< f_{B_{\perp}}^2 \right> 2\pi k_{\text{DM}} \Delta\text{DM} \left( \frac{1}{f_{\text{low}}^3} - \frac{1}{f_{\text{high}}^3} \right) \lesssim 2\pi\\
\frac{B_{\perp}}{1\,\text{G}} \lesssim \frac{1}{2.8} \left[ k_{\text{DM}} \Delta\text{DM} \left( \frac{1}{f_{\text{low}}^3} - \frac{1}{f_{\text{high}}^3} \right) \right]^{-1/2}.
\end{eqnarray}
Taking values corresponding to the wide-band Parkes Telescope observation, $\Delta\text{DM} \sim 0.1$\,pc\,cm$^{-3}$, $f_{\text{low}} = 705$\,MHz and $f_{\text{high}} = 4032$\,MHz, we find $B_{\perp} \lesssim 0.3$\,G in the eclipse region. Note, however, that this calculation is only valid for a (near-)perpendicular field and the limit could be larger should the magnetic field be directed away from the perpendicular to the line of sight, i.e. if there is a parallel component of similar or larger magnitude, which appears possible given the detected depolarisation of $L$.

\subsubsection{Faraday delay}\label{sec:faraday_delay}
\begin{figure}
	\includegraphics[width=\columnwidth]{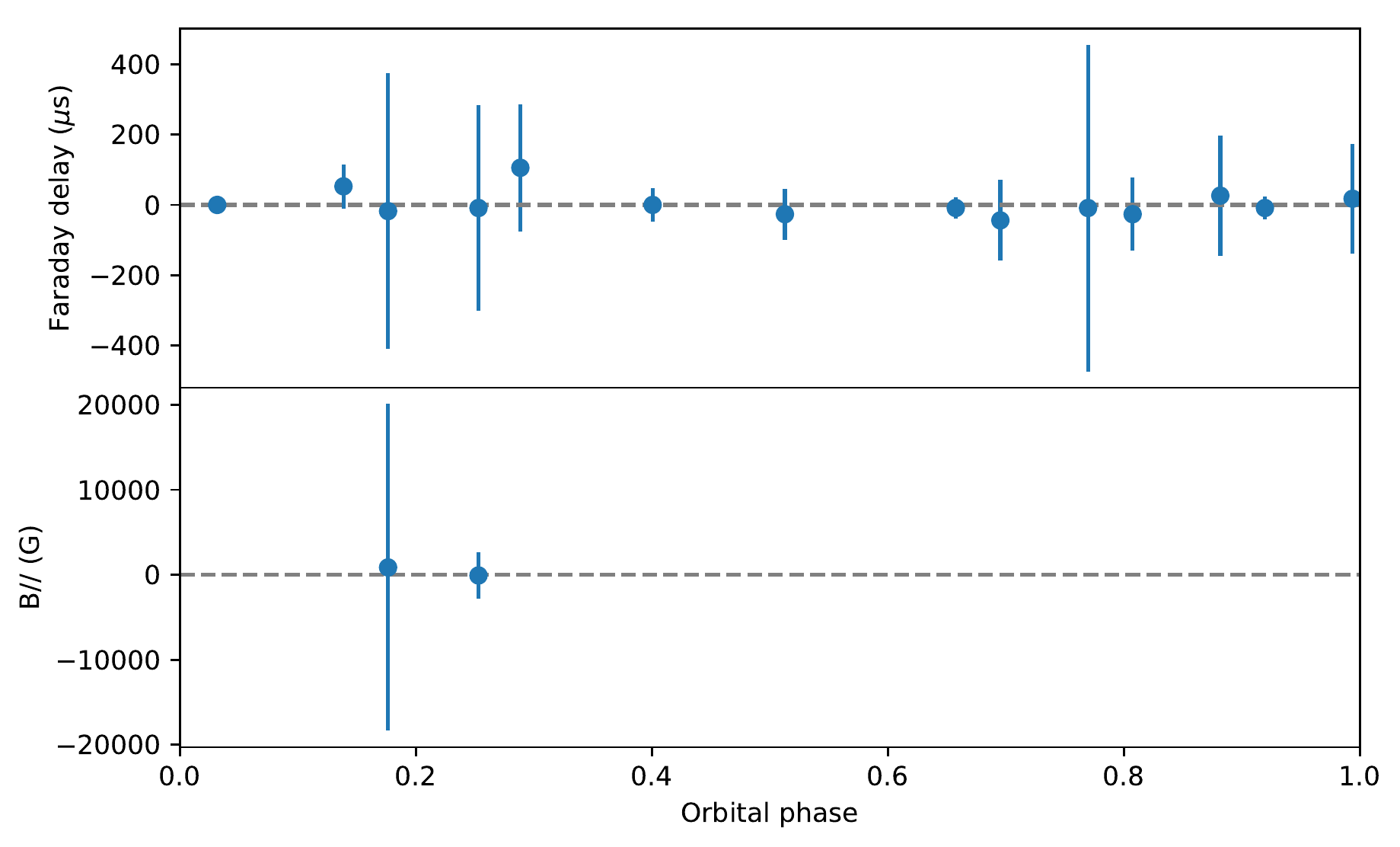}
	\caption[Faraday delays and implied magnetic field strengths corresponding to a Parkes Telescope observation of PSR J2051$-$0827 on 2018-Sep-23]{\textit{Top}: Measured Faraday delays between left- and right-hand circular polarisation profiles integrated over 16\,mins and 1241--1497\,MHz. \textit{Bottom}: The implied average magnetic field parallel to the line of sight, shown only for those points where $\Delta\text{DM} > 0$.}
	\label{fig:faraday_delay_dfb}
\end{figure}

\begin{figure}
	\includegraphics[width=\columnwidth]{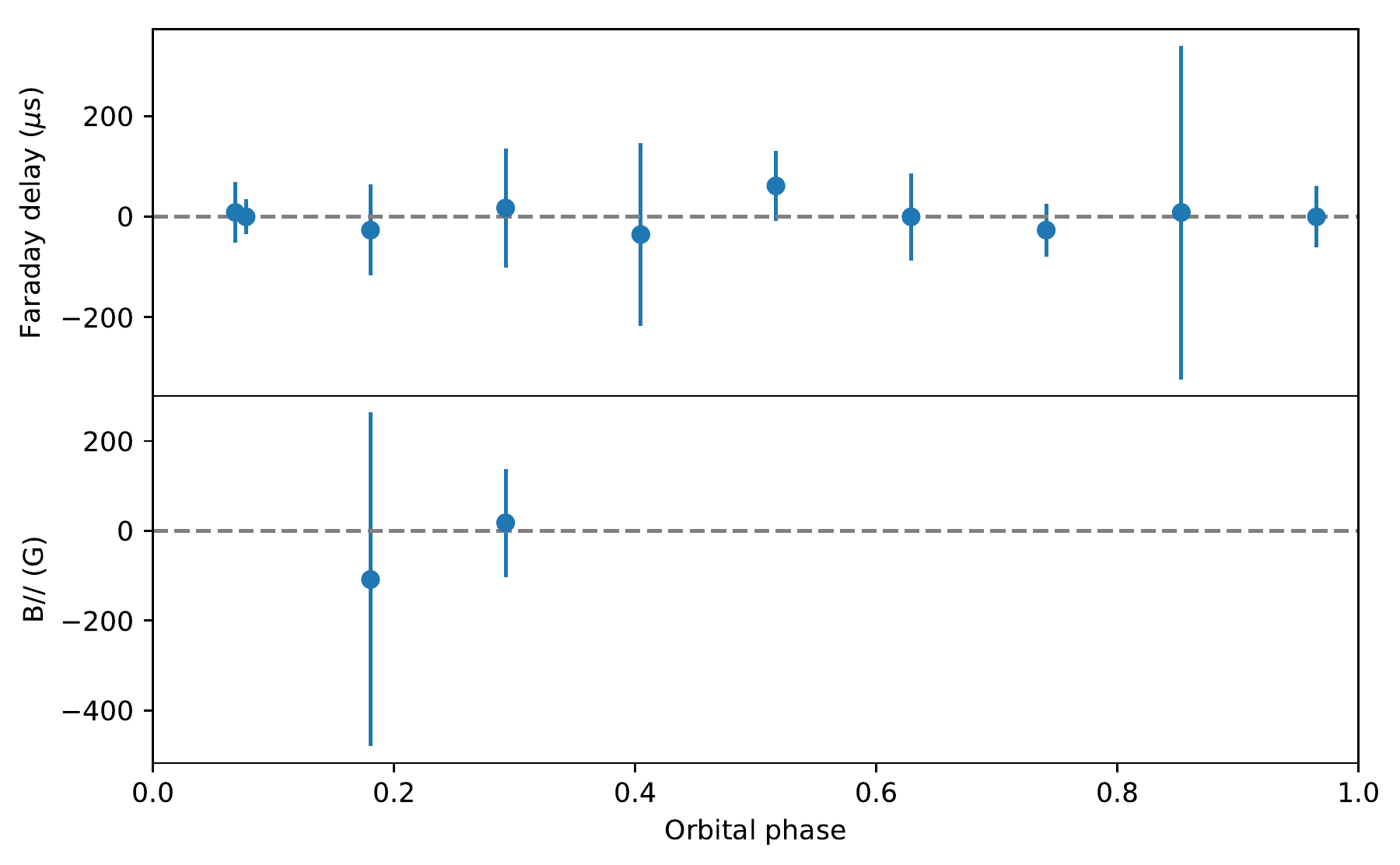}
	\caption[Faraday delays and implied magnetic field strengths corresponding to a Parkes Telescope observation of PSR J2051$-$0827 on 2018-Dec-03]{\textit{Top}: Measured Faraday delays between left- and right-hand circular polarisation profiles integrated over 16\,mins and 705--1345\,MHz. \textit{Bottom}: The implied average magnetic field parallel to the line of sight, shown only for those points where $\Delta\text{DM} > 0$.}
	\label{fig:faraday_delay_uwl}
\end{figure}

A final investigation of the polarisation properties of the eclipse concerns \textit{Faraday delay}. As considered in \citet{fbb+90} and \citet{llm+19} for PSR B1957+20, Faraday delay refers to the differential group delay between the left- and right-handed circular polarisation components which could become detectable in the presence of sufficiently large parallel magnetic fields. Such a study is useful here in the region where $L$ becomes depolarised, and no RM is measurable, while $V$ remains detectable.\\
The Faraday delay is expected to arise for propagation through a magnetised plasma and is given by,
\begin{equation}
\Delta t_{\text{FD}} = \frac{4 f_{B_{||}}}{f} \Delta t_f
\end{equation}
where $\Delta t_f$ is the excess delay induced by dispersion at frequency $f$ and $f_{B_{||}} = \frac{qB_{||}}{2\pi m_e} \approx 2.8 \left( \frac{B_{||}}{1\,\text{G}}\right)$\,MHz is the cyclotron frequency in terms of the average parallel component of the field \citep{llm+19}. Rearranging this we find,
\begin{equation}\label{eq:faraday_delay}
\frac{B_{||}}{1\,\text{G}} \approx \frac{1}{2.8} \frac{f}{4} \frac{\Delta t_{\text{FD}}}{\Delta t_f}.
\end{equation}
Using the full bandwidth of the first Parkes Telescope observation, and the low-frequency band (705--1345\,MHz) of the second Parkes Telescope observation -- where the delays would be expected to be the largest -- we cross-correlated the left- and right-hand circular polarisation profiles in integrations of 16\,mins with the template profile used previously in the flux fitting. Here we take positive $V$ to represent left-hand circularly polarised radiation, and negative $V$ to represent right-hand circularly polarised radiation, as defined as the PSR convention \citep{vmj+10}. The resulting cross-correlation functions are shown in Appendix C of the supplementary online material. For each integration, the difference between the maximum points of the left- and right-hand cross-correlation functions is taken as the Faraday delay. As in \citet{llm+19}, with the lack of an analytical method we estimated the uncertainties by bootstrapping. Here, this consisted of `resampling' each circular polarisation profile by replacing the flux in each profile bin with a value sampled from a normal distribution, $\mathcal{N}(\upmu,\sigma^2)$, with $\upmu$ equal to the original bin flux, and $\sigma$ equal to the sample standard deviation of the off-pulse bin fluxes in the original profile. The resampled profile was cross-correlated, and the difference between the maxima measured. This was repeated 1000 times for each profile, and the standard deviation of the 1000 delays was taken to be the uncertainty. The dispersive delay was estimated using $\Delta t_f \approx 4.15 \times 10^3 \frac{\Delta\text{DM}}{\nu_{\text{MHz}}^2}$\,seconds \citep{lk04}, with the mean $\Delta$DM within the integration and centre frequencies of 1369 and 1024\,MHz for the first and second observations, respectively.\\
The resulting Faraday delays are shown in Figs.~\ref{fig:faraday_delay_dfb} and \ref{fig:faraday_delay_uwl} along with the corresponding estimates of $B_{||}$ for those points with $\Delta t_f > 0$. The plots show that all of the $B_{||}$ values are consistent with zero, however the uncertainties are very large -- a consequence of both small dispersion delays at these relatively high observing frequencies, and low signal-to-noise of the circular polarisation profiles. The tightest constraint, where the dispersion delays are large approaching the eclipse egress in observation 2, suggests an average parallel field of $B_{||} = (20 \pm 120)$\,G. For a parallel magnetic field of the order of 100\,G, or less, in the eclipse region, assuming the orbital parameters stated previously, implies a magnetic field strength at the companion surface of $B_{\text{S}} = B_{||} \left(\frac{0.83}{0.15}\right)^3 \lesssim 2 \times 10^4$\,G. This is of the order of the surface field strengths considered previously for the companion in PSR J2051$-$0827 \citep{kmg00}.\\
\noindent
\noindent
Similarly to the Faraday rotation study above, this analysis suffers from the low signal-to-noise of the polarised radiation, requiring integrations of $\sim16$\,mins in order to achieve reliable cross-correlation spectra. These integrations are of the order of the eclipse duration, and as such average over the likely variable polarisation effects. Therefore, the Faraday delay measurements here are not sensitive to the case of a highly variable magnetic field, with a mean over the eclipse region of near to zero. Thus, the constraint here represents the average parallel field over the 16\,min integration, but does not directly rule out large variability. On the other hand, should fluctuations in the field between the different lines-of-sight be extreme then they would significantly broaden the integrated circular polarisation profiles as a result of averaging over differential delays. Fluctuations of the magnetic field of a similar order of magnitude to our uncertainties on $B_{||}$ would go undetected in this analysis.

\section{Eclipse mechanisms}\label{sec:mechanisms}
\citet{sbl+01} presented an investigation of possible eclipse mechanisms for the pulsed radio emission from PSR J2051$-$0827 in the frequency range 234--1660\,MHz. The study dismissed the possibility of pulse smearing (430\,MHz and above), refraction, reflection, free-free absorption and induced Compton scattering as possible eclipse mechanisms. However, the data available were not sufficient to be able to rule out either scattering of the pulses, pulse smearing at the lowest frequencies or cyclotron damping as the cause of the observed eclipses.\\
The observations that we present in this paper are consistent with the conclusions to dismiss the above-listed mechanisms, and here we further investigate those mechanisms deemed possible by the previous study, in light of the new data.

\subsection{Dispersion smearing}\label{sec:dm_smear}
\begin{figure}
	\includegraphics[width=\columnwidth]{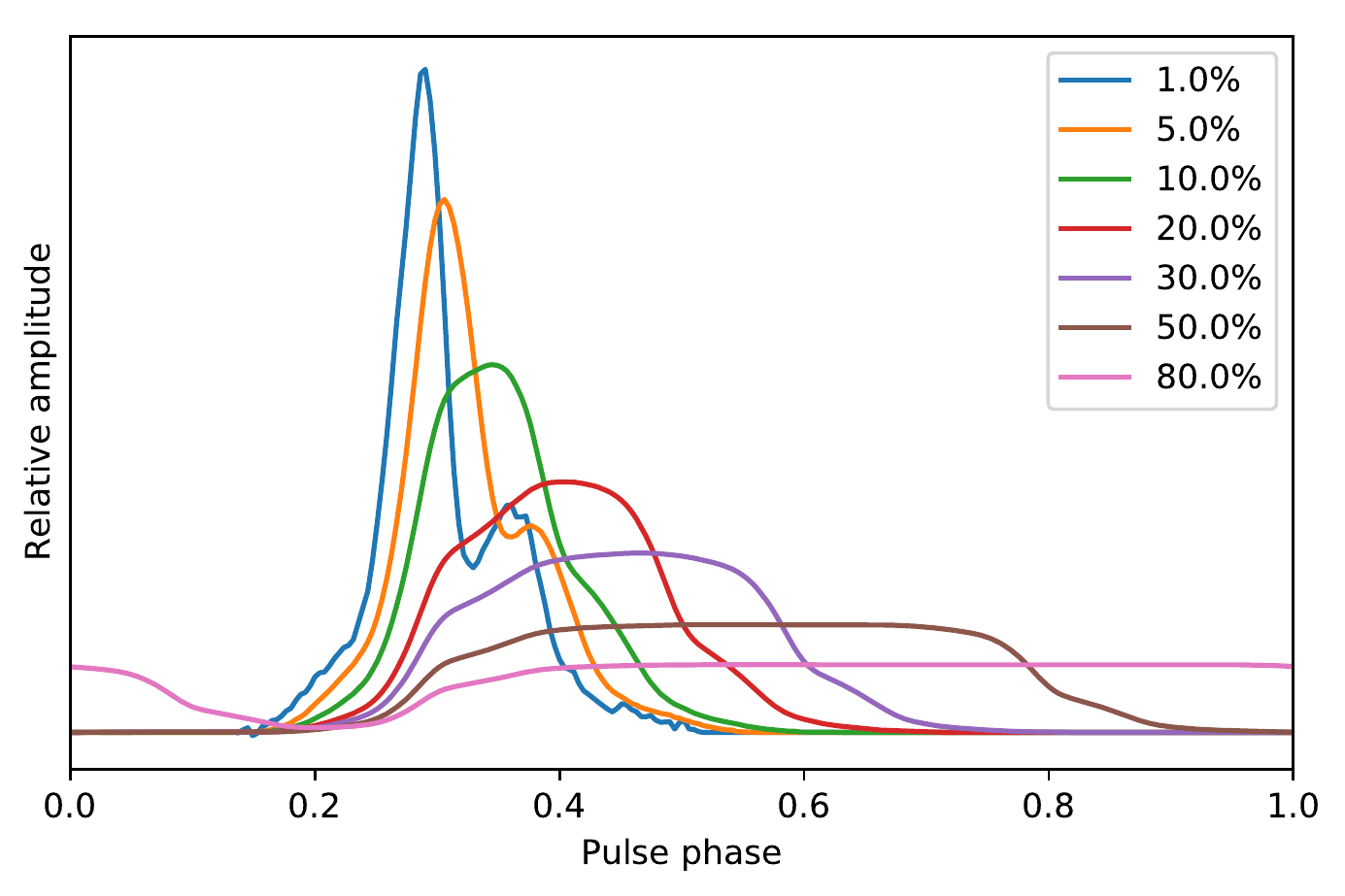}
	\caption[Modelled effect of dispersion smearing on pulse profiles of PSR J2051$-$0827]{Modelled effect of dispersion smearing on the 345\,MHz pulse profile. The model applies a linearly increasing dispersion measure delay to an array of pulses, then sums these to find the resulting integrated pulse profile. The smearing \% corresponds to the ratio: $100 \times (\text{dispersion delay between first and last pulse}) / (\text{pulse period})$.}
	\label{fig:dm_smear}
\end{figure}

\begin{figure*}
	\includegraphics[width=\textwidth]{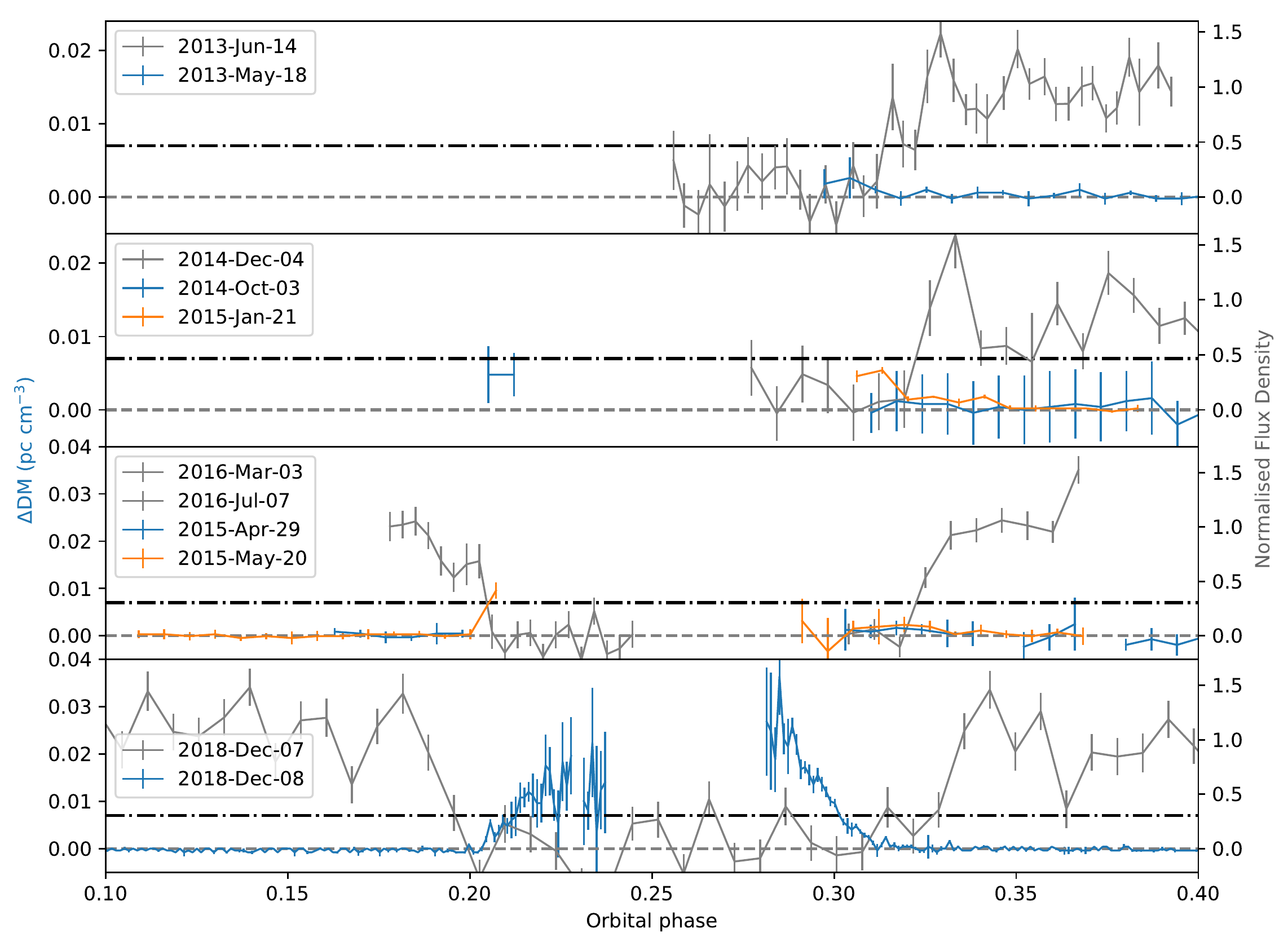}
	\caption[Flux density and DM variations in separate bands and epochs for PSR J2051$-$0827]{149\,MHz flux density (grey, solid) at eclipse edges, with the four panels showing different epochs to reduce the effect of temporal variability of the eclipses. Over-plotted are the deviations in dispersion measure from the out-of-eclipse level, measured from 345\,MHz observations (blue and orange). The dashed grey line simultaneously marks the out-of-eclipse dispersion level, and the minimum flux density detection limit of the telescope. The black dash-dotted line marks the dispersion smearing threshold required to broaden the 149\,MHz pulse by 30\% of the pulse period (see main text).}
	\label{fig:dm_smear_low}
\end{figure*}

\citet{sbl+01} concluded that smearing of the pulses, due to variations in DM within the typical integration times, was not sufficient to broaden the pulses beyond the intrinsic pulse width at frequencies of 430\,MHz and above. Their arguments could not, however, rule this out at lower-frequencies as the dispersion broadening increases as the inverse-square of observing frequency.\\
Fig.~\ref{fig:dm_smear} shows the modelled effect of dispersion smearing on the 345\,MHz pulse, where each smeared pulse represents an integration over a range of pulses delayed such as to model a linear change in DM over the integration time. Taking smearing of 30\% -- where the dispersion delay at the end of an integration is $0.3P$ larger than that at the beginning of the integration -- as sufficient to cause significant difficulty in detecting the pulse, the required change in DM to surpass this limit at 345\,MHz is $\sim0.04$\,pc\,cm$^{-3}$. Fig.~\ref{fig:Lband_dms} shows the measured DM variations throughout the eclipse region. In the most active eclipses (2011--2014) the DM changes over typical integration times are large enough to reach the required smearing limit, however often the DMs are not consistently this variable throughout the full eclipse region. This, in combination with the DM trends in the more quiescent periods (pre-2011, post-2014) where the variability is rarely sufficient to reach the required limit, shows that DM smearing cannot offer a reliable eclipse mechanism at 345\,MHz.\\
Similar calculations for the 149\,MHz radio emission yield a required change in DM, within a sub-integration duration, of only $\sim0.007$\,pc\,cm$^{-3}$. This is the same order of magnitude as the $1\sigma$ uncertainties in the high-frequency derived DMs, thus even for those observations where little deviation is seen we cannot rule out the possibility of the limit being surpassed throughout the bulk of the eclipse region. However, tighter constraints on the DM variations are provided by the 345\,MHz observations at the eclipse edges. Fig.~\ref{fig:dm_smear_low} shows the 149\,MHz flux density over-plotted with 345\,MHz DMs at four separate epochs, in order to suppress the effect of long-term eclipse variations. In all of the epochs the 149\,MHz flux drops into eclipse without the DM rising above the DM smearing limit calculated above. The only exception to this is the eclipse ingress in the third panel from the top, however even here the flux density begins to attenuate without any detected rise in DM. This provides significant evidence against DM smearing as a feasible eclipse mechanism, even at frequencies as low as 149\,MHz.
\noindent
\noindent

\subsection{Scattering}
In a similar fashion to DM smearing, scattering of the radio emission in the eclipse medium could act to broaden the pulses, and should the broadening become larger than the fundamental pulse width then the pulsations would become extremely difficult to detect. As with the above analysis, taking pulse broadening of 30\% of the pulse period as an approximate detection limit, for the 345\,MHz emission to become eclipsed would require a scattering timescale at 1400\,MHz of $\sim5\,\upmu$s ($0.001P$), assuming that the timescale scales as $\tau\propto\nu^{-4.4}$. This is smaller than the $1\sigma$ uncertainties on the scattering values measured from the L-band data, and often the scattering is seen to significantly exceed this during parts of the eclipse region, with $\Delta\tau \lesssim 1.1$\,ms\,($0.25P$) measured in the L-band observations in 2013--2014, coincident with the largest $\Delta$DMs. Similar calculations give a threshold scattering timescale at 345\,MHz of $\sim30\, \upmu$s ($0.007P$) in order for the 149\,MHz pulses to become eclipsed. This is only marginally greater than the typical $1\sigma$ uncertainties on the measured scattering values from the 345\,MHz data, and is occasionally measured to be exceeded at orbital phases corresponding to the LOFAR eclipse boundaries. This analysis favours broadening of the pulses through scattering as a possible eclipse mechanism, however higher signal-to-noise observations would be required in order to determine whether the scattering thresholds are always exceeded at the eclipse boundaries. It is worth noting that the scattering timescale is always measured to be $0 \lesssim \tau \lesssim 0.1P$ at the eclipse boundaries in both the 345\,MHz and 149\,MHz data, thus although there is often increased scattering present, it would need a fairly steep gradient in order to reach the $\sim0.3P$ threshold within a typical interval duration (20\,s -- 2\,min) at the eclipse boundaries.

\subsection{Cyclotron absorption}
Should magnetic fields be present in the eclipse region, then these can play a significant role in the mechanism behind the eclipses. The above analysis suggests that the average parallel component of the magnetic field in the typical eclipse region can not be much larger than $10^2$\,G. In addition, should the field be directed near-perpendicular to the line of sight, then the field is of the order of 1\,G or less. \citet{kmg00} specifically investigated the case of PSR J2051$-$0827; in their work they considered the companion to be a magnetic dwarf star, and suggested that eclipses occur as a result of cyclotron damping in a magnetosphere with a field strength at the companion surface of $10^4$--$10^7$\,G. With this model they derive a direct dependence between the frequency of damped radiation and the magnetic field strength in the eclipse region (Equation~13 in their paper):
\begin{equation}
\frac{\nu_d}{\text{1\,GHz}} \approx 2.8 \times 10^{-3} \frac{B_C}{\gamma_p (1-\cos\theta)} \left( \frac{R_C}{r} \right)^3,
\end{equation}
where $\nu_d$ is the frequency of radiation that is damped, $\gamma_p$ is the Lorentz factor of particles filling the companion magnetosphere, $\theta$ is the angle between the wave propagation and the magnetic field, and $B_C$, $R_C$ are the surface field strength in Gauss and radius of the companion star -- i.e. $B_C \left( \frac{R_C}{r} \right)^3$ represents the field strength in the eclipse region for a dipolar field. Taking our upper limit estimate of the eclipse field strength of $\sim10^2$\,G, and $\gamma_p(1-\cos\theta) \sim 1$, we find that frequencies of $\sim 300$\,MHz would be strongly damped, whereas those higher, i.e. $> 1$\,GHz, would be largely unaffected. This is in remarkable agreement with our observations. On the other hand, should the field strength be lower than this apparent upper limit, and Lorentz factors of 10--100 are considered as in \citet{kmg00}, then the model would predict that no eclipses would occur $> 100$\,MHz. In addition, their model relies on a magnetosphere filled with a pair plasma from the pulsar wind, which as discussed above appears unlikely given our detected depolarisation, although propagation through an unbalanced pair plasma could potentially satisfy both scenarios. In light of the significant assumptions that are required for both of our results, we deem this model to be still plausible, however further scrutiny of the model parameters and assumptions are now possible.\\
Prior to the development of this model, \citet{t+94} investigated the potential effects of cyclotron absorption in spider eclipses in terms of the optical depth at both the fundamental and higher harmonics as a function of the plasma properties. In contrast to \citet{kmg00}, this modelling does not assume a pair plasma, and in fact somewhat relies on Faraday rotation to ensure absorption of both assumed propagation modes. The authors find that for PSR B1957+20 an eclipse magnetic field of $\sim10$\,G, provided by either a companion magnetosphere or the magnetised pulsar wind, and a plasma temperature of $\sim 10^8$\,K can account for eclipses at frequencies $\lesssim 1$\,GHz, but higher temperatures would likely be required to absorb higher frequencies. Our observations are consistent with a magnetic field of similar strength in PSR J2051$-$0827, and as the authors note, such temperatures correspond to only a tiny fraction of the incident pulsar wind energy density, thus it appears plausible that this mechanism could indeed also account for the eclipses seen at low-frequency, and lack of eclipses at high-frequency, in PSR J2051$-$0827. We note that the same mechanism is favoured for eclipses in PSR J1810+1744 \citep{pbc+18}, PSR J2215+5135 \citep{bfb+16}, PSR J1227$-$4853 \citep{rrb+15} and PSR J1544+4937 \citep{brr+13}.

\section{Discussion}\label{sec: discuss}
\citet{wkh+12} infer the presence of an intra-binary shock between the pulsar and companion winds in the PSR J2051$-$0827 system from the observed X-ray spectra. Shock structure in the eclipse material could vastly increase the density compared to that implied by assuming the material is evenly distributed over a depth equivalent to the eclipse radius. \citet{asr+09} discussed the possibility of the density in the shock being high enough for the eclipses to be caused simply by the plasma frequency being larger than the observing frequency. From our observations the lowest frequency for which the radiation is not eclipsed is $\sim 700$\,MHz, and the tightest upper limit on the column density of material in the eclipse region is provided by the `quiescent' DM periods, in which the $1\sigma$ uncertainties suggest $N_e \lesssim 3 \times 10^{15}$\,cm$^{-2}$. In this case, for the plasma frequency to reach 700\,MHz would require a shock only $\sim5\times10^5$cm, or $\sim10^{-5}$ orbital radii thick, containing all of the intervening plasma column.\\
In modelling of intra-binary shocks in spider pulsars, \citet{whv+17} estimate that the contribution to magnetic fields in the shock region by the pulsar wind is of the order $0.1\,\text{G} \lesssim B \lesssim 200$\,G. This could feasibly account for the required magnetic fields for a number of theorised eclipse mechanisms -- e.g. cyclotron-synchrotron absorption \citep{t+94}, induced scattering \citep{lm95} -- and possibly the RM variations measured in this study, without the need to infer a companion magnetosphere. Alternatively, the vast extent of the low density ablated material that shocks the pulsar wind and the persistent structures that we observe it to form (Fig.~\ref{fig:Lband_dms}), can be attractively explained by a magnetic field provided by the companion star giving support to the material, as is also noted in \citet{s+96,sbl+01}. Further modelling of intra-binary shocks, presented in \citet{wvh+18}, suggests that pressure balance in the shock can be achieved in a number of redback pulsars if the companion has a surface magnetic field strength on the order of kilogauss. It is interesting to note that such a field is consistent with the loose constraints found here, and is also of a similar magnitude to that required for cyclotron damping to cause the eclipses \citep{kmg00}.\\
Such a magnetic field can be inferred to account for the orbital fluctuations that are detected in PSR J2051$-$0827 \citep{lvt+11,svf+16} due to the companion star's magnetic cycles causing variations in its oblateness \citep{as94}. These variations could presumably be expected to influence the ablation of the companion and consequently the eclipsing medium which forms. Comparison of the long-term changes in the high-frequency DM patterns, measured here, with the timing variations presented in \citet{svf+16} reveals no clear correlation between the two, however this is not surprising based on the short overlap in our datasets relative to the timescale of the orbital variations. We note that the rise and fall-off of eclipse DMs that we detect over our observation range could be interpreted as following a similar trend, albeit with a lag of $\sim500$\,days, as the lattermost peak in the measured time of ascending node parameter \cite[Fig.\,6 of][]{svf+16}, although more observations would be required to see if such a pattern persists. A more stringent test of this would be provided by comparison of the two measurements during a less `quiescent' time of the orbital parameter variations, or in a different eclipsing system with more pronounced orbital fluctuations. On the other hand, with the apparent lack of correlation between the measured high-frequency DMs and the low-frequency eclipse variations, it may be that the DMs that we detect are not very representative of the parameters of the eclipse medium as a whole, especially when the line of sight is so far inclined from the orbital plane.\\
Inferring a mass loss rate for the companion of PSR J2051$-$0827 has an added complication due to the small $\sim40^{\circ}$ inclination angle of the orbit, suggesting that our line of sight samples only the outer edges of the eclipse medium. Somewhat naively we can make an estimate by assuming that the material is spherically symmetric about the companion, in which case the material is ejected from the system through a circular region, centred on the companion, perpendicular to the orbital plane. With this assumption we can follow the analysis in Section~6.3 of \citet{t+94}, whereby the ablated material, approximated to balance the momentum flux of the pulsar wind at the orbital separation, becomes entrained in the pulsar wind. Given the orbital parameters listed in Section~\ref{sec: intro}, and taking the eclipse duration to be 10\% of the orbit, the corresponding eclipse width is $\sim0.7\,R_{\odot}$. As this corresponds to a chord across the projected circle of material around the companion due to our inclined view of the system, we estimate the actual diameter of the circle to be $\sim1.8\,R_{\odot}$. Taking the column density of material to be $\sim0.1$\,pc\,cm$^{-3}$ (note that this could be much higher in the orbital plane, closer to the companion), and the depth to be of the order of the eclipse radius, the density of material in this region is $n_{\text{e}} \sim10^7$\,cm$^{-3}$. The velocity of the material entrained in the pulsar wind is then $V_{\text{W}} = (U_{\text{E}}/n_{\text{e}}m_{\text{p}})^{1/2} \sim 5\times10^8$\,cm\,s$^{-1}$, where $U_{\text{E}} \sim 2.4$\,ergs\,cm$^{-3}$ is the energy density of the pulsar wind at the companion. This leads to an estimated mass loss rate of $\dot{M}_{\text{C}} \sim \pi R_{\text{E}}^2m_{\text{p}}n_{\text{e}}V_{\text{W}} \sim 10^{-12}\,M_\odot$\,yr$^{-1}$; similar to that calculated for black widow PSR J1810+1744 \citep{pbc+18} and a few times that of PSR B1957+20 \citep{t+94}. This value, although subject to much uncertainty, is likely to act more as an upper limit on the long-term evolution of the system as the orbital separation is expected to increase as mass is lost from the system, reducing the irradiative influence of the pulsar wind. In addition, the presence of a companion magnetosphere could effectively contain, or shield the ablated material, reducing the estimated rate of mass loss from the system. With these caveats, the observations presented here appear to disfavour the possibility of the companion being fully evaporated within a Hubble time. On the other hand, should the ionisation fraction of the ablated material be low then the DM measurements -- sensitive only to ionised electrons -- underestimate the true column density of material and as such could result in a larger mass loss rate. We note that a large neutral component may be difficult to achieve, given the close proximity of the material to the intense pulsar wind.

\subsection{Mini-eclipses}
\begin{figure}
	\includegraphics[width=\columnwidth]{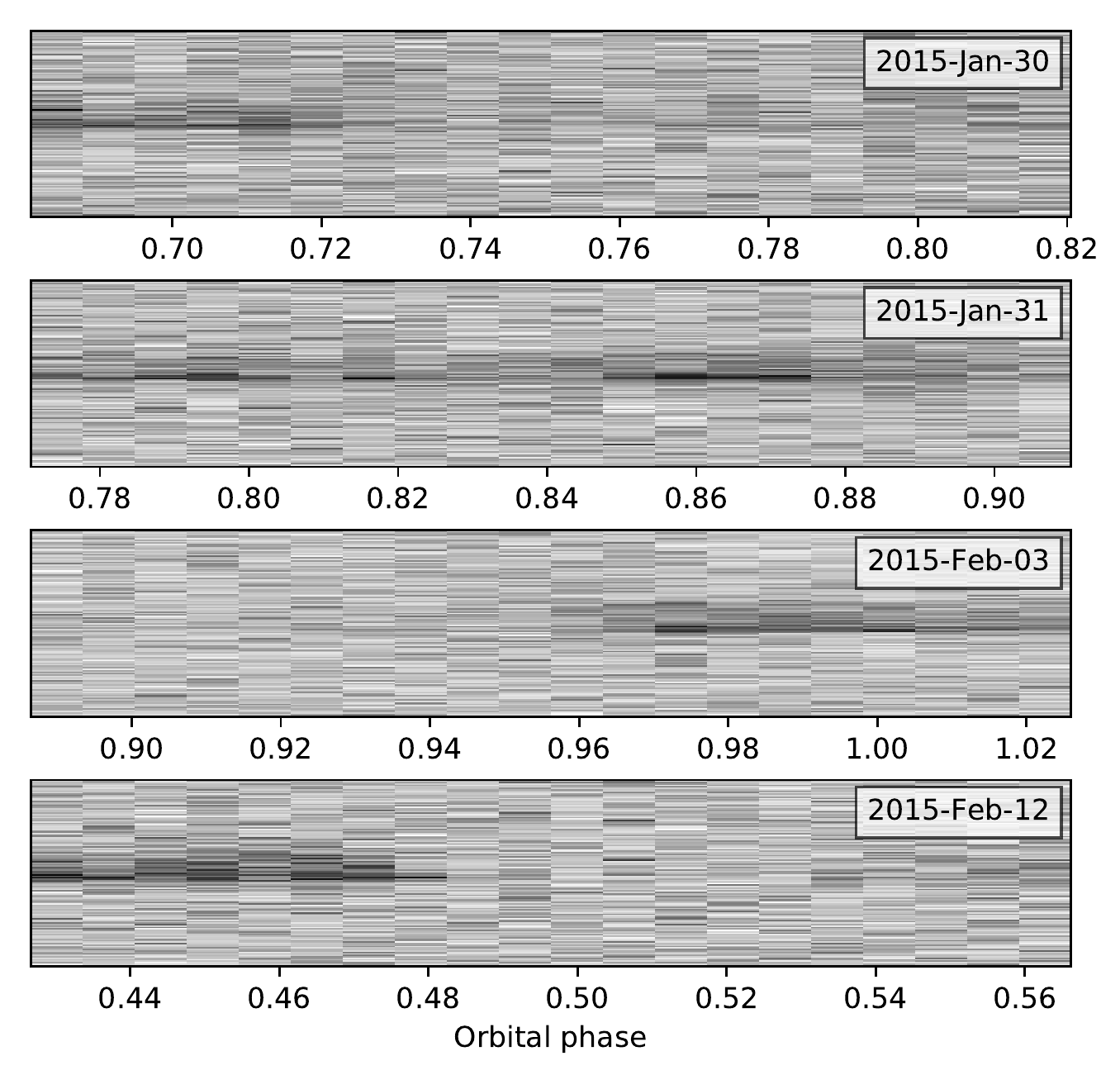}
	\caption[Short duration flux density variations in PSR J2051$-$0827 at 149\,MHz]{Short duration flux density modulations detected in 149\,MHz observations over a 2\,week period in 2015. The modulations occur at orbital phases far from the typical eclipse.}
	\label{fig:mini_ecl}
\end{figure}

\begin{figure}
	\includegraphics[width=\columnwidth]{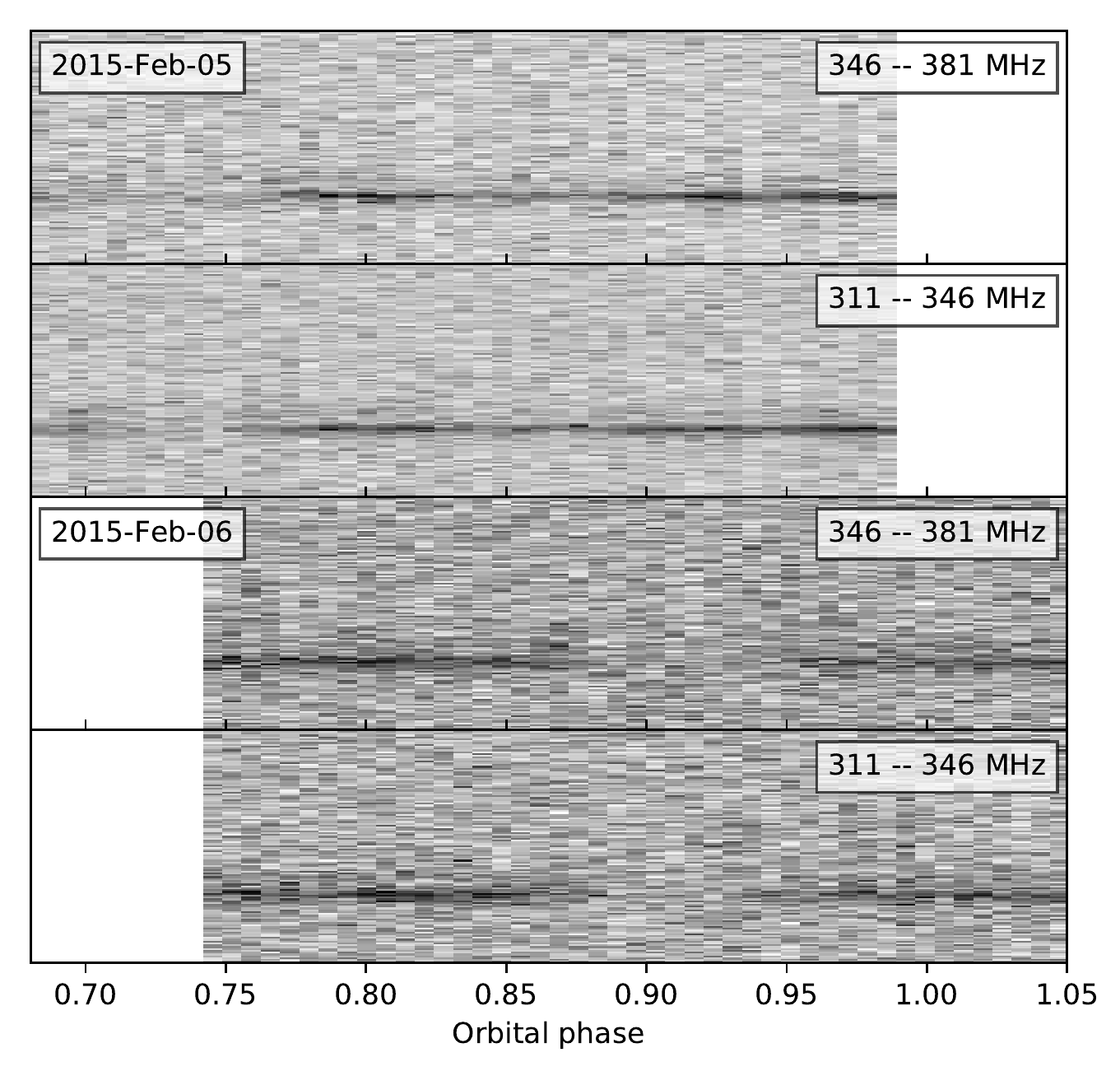}
	\caption[Short duration flux density variations in PSR J2051$-$0827 at 345\,MHz]{Short duration flux density modulations detected in 345\,MHz WSRT observations on 2 consecutive days in 2015. Each observation has been split over 2 panels, showing the top and bottom halves of the 70\,MHz bandwidth to highlight the wide band nature of the flux density variations, distinguishing these from diffractive scintillation (see main text).}
	\label{fig:mini_ecl_05}
\end{figure}

Figs.~\ref{fig:mini_ecl} and \ref{fig:mini_ecl_05} show short duration flux density reductions -- `mini-eclipses' -- at orbital phases far from the main eclipse at 149\,MHz and 345\,MHz, respectively. Such mini-eclipses are more commonly associated with redback systems that have much larger main eclipse fractions \citep{lmd+90,asr+09,drc+16}. It is interesting to note that all of the mini-eclipses were detected in a 2\,week period in 2015-Feb, and no other convincing mini-eclipses were seen at any other time. The small diffractive interstellar scintillation (DISS) bandwidths of $\sim1$\,kHz and $\sim50$\,kHz, for 149\,MHz and 345\,MHz centre frequencies respectively, for PSR J2051$-$0827 mean that any effects of DISS will be well averaged out over the observing bandwidths, and thus cannot be responsible for such variations. These are of particular interest for eclipse mechanisms that require the presence of a companion magnetosphere, as this would have no influence at orbital phases so far from companion inferior conjunction. Considering the lack of magnetic fields to shield this material from the pulsar wind, it presumably must have been ejected far from the orbit in the time taken to cross our line of sight. No significant pulse delay or broadening is seen around the mini-eclipses, suggesting that the material causing the reduction in flux density either has a very low density, or relatively sharp boundaries such that the increase and decrease in density evades detection within the observation time integrations. The presence of material at multiple orbital phases, all observed within a short 2\,week period may be linked to variations in the intra-binary shock or flaring of the companion star, such as those suggested by \citet{chb18} and \citet{ylk+18}, which could lead to more erratic mass ejection from the system. Alternatively, the balance between the material outflow and the pulsar wind may be such that the material can actually fall in towards the pulsar, leading to variable eclipse fractions at different orbital phases; this is the scenario suggested for eclipses in Ter5A \citep{tb93}.

\subsubsection{Effects of Earth's ionosphere}
A recent study by Scholte et al. (in prep.) into the effects of the ionosphere on LOFAR observations of pulsars found correlations between the measured pulsar flux density in beamformed data with the variable apparent position of sources in the image-plane. The variations were seen to occur on timescales of minutes and were attributed to turbulence in the ionosphere displacing the target source from the centre of the low-frequency tied-array beam, thus reducing its sensitivity. Although it is not clear how the effects of the turbulence scale with the line of sight TEC of the ionosphere, we note that the TEC values at times corresponding to the mini-eclipse observations, found from the \textsc{ionFR} code introduced in Section~\ref{sec: rm}, are a factor of $\gtrsim2$ larger than the median value for all of our observations ($\sim 2.5 \times 10^{13}$\,cm$^{-2}$). Since TEC values of a similar magnitude are seen in only a few of the other observations, we briefly investigate here the possibility of ionospheric turbulence being the cause of the apparent mini-eclipses.\\
In the LOFAR mini-eclipse observed on 2015-Feb-12 the flux density of the pulsar drops to $\lesssim 5$\% of the pre-eclipse value. By approximating the LOFAR tied-array beam as a 2D Gaussian, with full-width at half-maximum determined by $\lambda / D$ where $\lambda = 150$\,MHz and $D$ to be the maximum 2\,km baseline, we estimate the required offset from the centre of the beam to attenuate the sensitivity to 5\% to be $\sim3.5$\,arcmin. Equivalently, for the WSRT mini-eclipse on 2015-Feb-06 the pulsar flux density drops to $\sim20$\% of the pre-eclipse value. The first-order beam model for WSRT attenuates as cos$^6(c\nu r)$, where $c \approx 66$ at $\nu=345$\,MHz and $r$ is the distance from the beam centre in degrees\footnote{\href{http://www.astron.nl/radio-observatory/astronomers/wsrt-guide-observations/wsrt-guide-mffe-observations}{http://www.astron.nl/radio-observatory/astronomers/wsrt-guide-observations/wsrt-guide-mffe-observations} ; Section~5.7, accessed 2019-03.}. Thus, for the sensitivity to attenuate to 20\% requires a source offset of $\sim1.8$\,arcmin. For the 8, 30\,min observations analysed in Scholte et al. (in prep.), the maximum deviation of any source in the field of view was $\sim2$\,arcmin and the maximum deviation averaged over individual 30\,min observations was $\sim0.6$\,arcmin. Although the maximum observed offset is a similar order of magnitude to our required values, the duration spent in such a large offset was $\lesssim 1$\,min; far from the $\sim5-10$\,min duration of the mini-eclipses. In addition, if the angular offsets were to scale as $\propto \nu^{-2}$ as expected from Kolmogorov turbulence, then the required offset for the 345\,MHz mini-eclipse would correspond to an offset of $\sim10$\,arcmin at LOFAR frequencies. To our knowledge, deviations so large have never been seen in LOFAR image-plane observations.\\
Furthermore, the above calculations assume symmetric beams with projected baselines equivalent to the physical baseline lengths; an approximation only valid for observations at the zenith angle. At beam pointing angles closer to the horizon, applicable to the observations here, the projected baseline lengths reduce and result in a wider, asymmetric beam. In this case the angular offsets required to cause the suggested sensitivity reductions would be even larger than those calculated above. Thus, although ionospheric effects may cause detectable variation in the pulsar flux density, it does not appear sufficient to account for the observed mini-eclipses, especially at 345\,MHz.
\noindent
\noindent

\section{Conclusions}
We present a large number of observations of the black widow pulsar PSR J2051$-$0827 at observing frequencies spanning 149--4032\,MHz, over $\sim10$\,years. Fitting for dispersion and scattering delays simultaneously, we detect variations in the material ablated from the companion on timescales ranging from shorter than the 2.4\,hr orbital period, to multiple years. These pronounced fluctuations in the DMs throughout the eclipse region do not show any clear correlation with the variable eclipse widths seen at lower observing frequencies, nor do they convincingly relate to the timing variations presented in \citet{svf+16}. On this topic, we plan to continue to monitor the source with regular 1\,hr observations using the Lovell Telescope, allowing a thorough test of whether or not a pattern develops between the eclipse and timing behaviours. The current lack of correlation may be due to the variable DMs not being representative of the behaviour of the eclipse material as a whole, which would be understandable bearing in mind the $\sim40^{\circ}$ inclination of the orbit.\\
We detect DM structures that persist for many months, present a possible increase in variance in Faraday rotation values in eclipse egress, and find depolarisation of the linear polarisation components of the pulse throughout the low-frequency eclipse region, all of which could be explained by magnetic fields in the ablated material. Consideration of the depolarisation of the linear components, and Faraday delay measurements between the two hands of circular polarisation, allow us to place limits on the average magnetic field parallel to the line of sight of $10^{-4}\,\text{G} \lesssim B_{||} \lesssim 10^2$\,G in the eclipse region. In addition, if the field were to be directed near-perpendicular to the line of sight, then a lack of depolarisation of the circular polarisation components allows us to constrain $B_{\perp} \lesssim 0.3$\,G. These are the first available limits for an eclipse region magnetic field in PSR J2051$-$0827. We argue that such fields are consistent with those required for cyclotron damping \citep{kmg00,t+94} as an eclipse mechanism, although scattering of the low-frequency emission is also fully consistent with the observations, in agreement with the calculations of \citet{sbl+01}.\\
An updated estimate of the mass loss rate from the companion, $\dot{M}_{\text{C}} \sim 10^{-12}\,M_\odot$\,yr$^{-1}$, is orders of magnitude higher than previously calculated in \citet{s+96}, although still appears to be too low to realistically fully ablate the companion in a Hubble time. In addition, we show a series of mini-eclipses detected over a 2\,week period at orbital phases far from inferior conjunction of the companion. Such a phenomena is more commonly associated with redback systems, and any models of these systems would need to infer where such material is located along the line of sight.

\section*{Acknowledgements}
EJP would like to thank Simon Johnston and James Green for helping with planning and observing with the Parkes Telescope, and Andrew Lyne and Mitchell Mickaliger for their work on recent, dedicated Lovell observations.\\
This paper is based (in part) on data obtained with the International LOFAR Telescope (ILT). LOFAR is the Low Frequency Array designed and constructed by ASTRON. It has observing, data processing, and data storage facilities in several countries, that are owned by various parties (each with their own funding sources), and that are collectively operated by the ILT foundation under a joint scientific policy. The ILT resources have benefitted from the following recent major funding sources: CNRS-INSU, Observatoire de Paris and Universit\'{e} d'Orl\'{e}ans, France; BMBF, MIWF-NRW, MPG, Germany; Science Foundation Ireland (SFI), Department of Business, Enterprise and Innovation (DBEI), Ireland; NWO, The Netherlands; The Science and Technology Facilities Council, UK; Ministry of Science and Higher Education, Poland.\\
The Westerbork Synthesis Radio Telescope is operated by the ASTRON (Netherlands Institute for Radio Astronomy) with support from the Netherlands Foundation for Scientific Research (NWO).\\
The Parkes radio telescope is part of the Australia Telescope which is funded by the Commonwealth of Australia for operation as a National Facility managed by CSIRO.\\
Pulsar research at Jodrell Bank Centre for Astrophysics and Jodrell Bank Observatory is supported by a consolidated grant from the UK Science and Technology Facilities Council (STFC).\\
We thank the staff of the GMRT that made the observations possible. GMRT is run by the National Centre for Radio Astrophysics of the Tata Institute of Fundamental Research.\\
EJP acknowledges support from a UK Science and Technology Facilities Council studentship. RPB acknowledges support from the ERC under the European Union's Horizon 2020 research and innovation programme (grant agreement No. 715051; Spiders). SO acknowledges Australian Research Council grant FL150100148.




\bibliographystyle{mnras}
\bibliography{../../Documents/bibliography4} 






\bsp	
\label{lastpage}
\end{document}